\documentclass[10pt]{article}

\usepackage{epsfig}
\usepackage{latexsym}
\usepackage{amssymb}
\usepackage{amsmath}
\usepackage{proof}
\usepackage{pstricks}
\usepackage{times,latex8}
\usepackage{mathrsfs}
\usepackage{subfigure}
\usepackage{isolatin1}
\usepackage{url}

\newcommand{\F}{\mathbb{F}}

\newcommand{\N}{\mathbb{N}}

\newcommand{\D}{\mathbb{D}}

\newcommand{\h}[1]{\mathsf{H}(#1)}
\newcommand{\rh}[1]{\mathsf{RH}(#1)}

\newcommand{\llan}{\langle\!\langle}
\newcommand{\rran}{\rangle\!\rangle}
\newcommand{\llcu}{\{\hspace{-.9mm}\{}
\newcommand{\rrcu}{\}\hspace{-.9mm}\}}

\newcommand{\ca}[2]{\mathcal{R}_{#1}(#2)}

\newtheorem{lemma}{Lemma}
\newtheorem{proposition}{Proposition}
\newtheorem{theorem}{Theorem}

\newenvironment{proof}{\begin{trivlist}
       \item[\hskip \labelsep {\bfseries Proof.}]}{\end{trivlist}}

\newenvironment{lemma*}[1][Lemma]{\begin{trivlist}
       \item[\hskip \labelsep {\bfseries Lemma #1}]\it}{\end{trivlist}}
\newenvironment{proposition*}[1][Proposition]{\begin{trivlist}
       \item[\hskip \labelsep {\bfseries Proposition #1}]\it}{\end{trivlist}}
\newenvironment{theorem*}[1][Theorem]{\begin{trivlist}
       \item[\hskip \labelsep {\bfseries Theorem #1}]\it}{\end{trivlist}}
   
\newcommand{\linear}{\multimap}
\newcommand{\linearo}[1]{\stackrel{#1}{\multimap}}

\newcommand{\Coerc}{\mathbf{Coerc}}

\newcommand{\Duplicate}{\mathbf{Duplicate}}

\newcommand{\Predecessor}{\mathbf{Predecessor}}

\newcommand{\Add}{\mathbf{Add}}
\newcommand{\Square}{\mathbf{Square}}
\newcommand{\Exp}{\mathbf{Exp}}
\newcommand{\Blowup}{\mathbf{Blowup}}
\newcommand{\Leaves}{\mathbf{Leaves}}
\newcommand{\Extract}{\mathbf{Extract}}
\newcommand{\inttou}[1]{\ulcorner #1\urcorner}
\newcommand{\bintob}[1]{\ulcorner #1\urcorner}

\newcommand{\spb}{\;|\;}
\newcommand{\qed}{\hfill$\Box$}

\newcommand{\A}{\mathbb{A}}
\newcommand{\B}{\mathbb{B}}
\newcommand{\C}{\mathbb{C}}
\newcommand{\G}{\mathbb{G}}

\newcommand{\U}{\mathbb{U}}

\newcounter{number}
\newenvironment{varitemize}
{
\begin{list}{\labelitemi}
{
\setlength{\itemsep}{0pt}
 \setlength{\topsep}{0pt}
 \setlength{\parsep}{0pt}
 \setlength{\partopsep}{0pt}
 \setlength{\leftmargin}{15pt}
 \setlength{\rightmargin}{0pt}
 \setlength{\itemindent}{0pt}
 \setlength{\labelsep}{5pt}
 \setlength{\labelwidth}{10pt}}}
{
 \end{list} 
}

\begin{document}
  \title{The Geometry of Linear Higher-Order Recursion\thanks{The 
      author is partially supported by PRIN projects
      PROTOCOLLO (2002) and FOLLIA (2004).}} 
  \author{Ugo Dal Lago\\
    {\it Dipartimento di Scienze dell'Informazione}\\ 
    {\it Universit\`a degli Studi di Bologna, Italy}\\
    \texttt{dallago@cs.unibo.it}}
\maketitle
\begin{abstract}
Linearity and ramification constraints have been widely used to
weaken higher-order (primitive) recursion in such a way that
the class of representable functions equals the class
of polytime functions, as the works by Leivant, Hofmann and others show. 
This paper shows that fine-tuning these two constraints leads to different expressive 
strengths, some of them lying well beyond 
polynomial time. This is done by introducing a new semantics, called
algebraic context semantics. The framework stems from Gonthier's
original work and turns out to be a versatile and powerful tool 
for the quantitative analysis of normalization in the lambda-calculus with
constants and higher-order recursion.
\end{abstract}
\section{Introduction}
Implicit computational complexity aims at giving machine-independent
characterizations of complexity classes. In recent years, the field has 
produced a number of interesting results. Many of them relate complexity
classes to function algebras, typed lambda calculi and logics by
introducing appropriate restrictions to (higher-order) primitive recursion
or second-order linear logic. The resulting subsystems are then shown
to correspond to complexity classes by way of a number of different,
heterogeneous techniques. Many kinds of constraints have been
shown to be useful in this context; this includes 
ramification~\cite{Bellantoni92CC,Leivant93popl,Leivant99apal}, linear
types~\cite{hofmann00safe,Bellantoni00apal,Leivant99csl,dallago03types} and restricted 
exponentials~\cite{Girard98ic,Lafont04tcs}. 
However, the situation is far from being satisfactory. There are still 
many open problems: for example, it is not yet clear what the 
consequences of combining different constraints are. Moreover, using 
such systems as a foundation for
resource-aware programming languages relies heavily on them to be able
to capture interesting algorithms. Despite some recent 
progresses~\cite{Hofmann99lics,Bonfante04tcs}, a lot of work still has 
to be done.

Undoubtedly, what is still lacking in this field is a powerful and
simple mathematical framework for the analysis of quantitative aspects 
of computation. Indeed, existing systems have been often studied 
using ad-hoc techinques which cannot be easily adapted
to other systems. A unifying framework would not just make the task
of proving correspondences between systems and complexity classes
simpler, but could be possibly used \emph{itself} as a basis for
introducing resource-consciousness into programming languages.
We believe that ideal candidates to pursue these goals are Girard's 
geometry of interaction~\cite{Girard89lc,Girard88cl} and related frameworks, 
such as context semantics~\cite{Gonthier92popl,Mairson02fsttcs}. 
Using the above techniques as tools in the study of complexity of normalization
has already been done by Baillot and Pedicini in the context of elementary
linear logic~\cite{baillot01fi}, while
game models being fully abstract with respect to 
operational theory of improvement~\cite{Sands91} have recently been proposed 
by Ghica~\cite{ghica05popl}. Ordinal analysis has already been proved useful to
the study of ramified systems (e.g.~\cite{Wainer02,Simmons05jsl}) but, to the author's knowledge,
the underlying framework has not been applied to linear calculi.
Similarly, Leivant's instrinsic reasoning framework~\cite{Leivant02apal,Leivant04tcs} can help defining and
studying restrictions on first-order arithmetic inducing complexity bounds on provably 
total functions: however, the consequences of linearity conditions cannot 
be easily captured and studied in the framework.

In this paper, we introduce a new 
semantical framework for higher-order recursion, called algebraic context semantics. It
is inspired by context semantics, but designed to be a tool for proving
\emph{quantitative} rather than qualitative properties of programs. As we will see, it turns
out to be of great help when analyzing quantitative aspects of normalization
in presence of linearity and ramification constraints.
Informally, algebraic context semantics allows to prove bounds on the
\emph{algebraic potential size} of System \textsf{T} terms, where the
algebraic potential size of any term $M$ is the maximum size of free algebra terms
which appear as subterms of reducts of $M$. As a preliminary result, the algebraic
potential size is shown to be a bound to normalization time, modulo a polynomial
overhead. Consequently, bounds obtained through context semantics translate
into bounds to normalization time.
 
Main results of this work are sharp characterizations of the expressive power of
various fragments of System \textsf{T}. Almost all of them are novel.
Noticeably, these results are obtained in a uniform way and,
as a consequence, most of the involved work has been factorized over
the subsystems and done just once. Moreover, we do not simply prove
that the class of representable first-order functions equals
complexity classes but, instead, we give bounds on the time
needed to normalize \emph{any} term. This makes our results
stronger than similar ones from the literature~\cite{hofmann00safe,Bellantoni00apal,Leivant99csl}. 
Our work gives some answers to a fundamental question implicitly raised by 
Hofmann~\cite{hofmann00safe}: 
are linearity conditions sufficient to keep the expressive power of
higher-order recursion equal to that of first-order recursion? In particular,
a positive answer can be given in case ramification
does not hold. The methodology introduced here can be applied to multiplicative
and exponential linear logic~\cite{dallago06lics}, 
allowing to reprove soundness results for various subsystems of the logic.

The rest of the paper is organized as follows: in Section~\ref{sect:syntax}
a call-by-value lambda calculus will be described as well as 
an operational semantics for it; in Section~\ref{sect:subsystems} 
we will define ramification and linearity conditions on the underlying type system,
together with subsystems induced by these constraints; in 
Section~\ref{sect:acs} we motivate and introduce algebraic
context semantics, while in Section~\ref{sect:cn} we will use
it to give bounds on the complexity of normalization.
Section~\ref{sect:completeness} is devoted to completeness
results.

\section{Syntax}\label{sect:syntax}
In this section, we will give some details on our reference system, namely a formulation
of G\"odel's \textsf{T} in the style of Matthes and Joachimsky~\cite{Matthes03aml}.
The definitions will will be standard. The only unusual aspect of our syntax is the
adoption of weak call-by-value reduction. This will help in keeping the language
of terms and the underlying type system simpler.

Data will be represented by terms in some free algebras. As it will be shown,
different free algebras do not necessarily behave in the same way
from a complexity viewpoint, as opposed to what happens in computability theory.
As a consequence, we cannot restrict ourselves to a canonical free algebra and need
to keep all of them in our framework.
A {\it free algebra} $\A$ is a couple $(\mathcal{C}_\A,\mathcal{R}_\A)$
where $\mathcal{C}_\A=\{c_1^\A,\ldots,c^\A_{k(\A)}\}$ is a finite set of {\it constructors} and
$\mathcal{R}_\A:\mathcal{C}_\A\rightarrow\N$ maps every constructor
to its {\it arity}. If the underlying free algebra $\A$ is clear from the context,
we simply write $\mathcal{R}(c)$ in place of $\mathcal{R}_\A(c)$.
A free algebra $\A=(\{c^\A_1,\ldots,c^\A_{k(\A)}\},\mathcal{R}_\A)$
is a \emph{word algebra} if
\begin{varitemize}
  \item
   $\mathcal{R}(c^\A_i)=0$ for one (and only one) $i\in\{1,\ldots,k(\A)\}$;
  \item
   $\mathcal{R}(c^\A_j)=1$ for every $j\neq i$ in $\{1,\ldots,k(\A)\}$.
\end{varitemize}
If $\A=(\{c^\A_1,\ldots,c^\A_{k(\A)}\},\mathcal{R}_\A)$ is a word algebra,
we will assume $c^\A_{k(\A)}$ to be the distinguished element of
$\mathcal{C}_\A$ whose arity is $0$ and $c^\A_1,\ldots,c^\A_{k(\A)-1}$ will denote
the elements of $\mathcal{C}_\A$ whose arity is $1$.
$\U=(\{c^\U_1,c^\U_{2}\},\mathcal{R}_\U)$ is 
the word algebra of unary strings.  
$\B=(\{c^\B_1,c^\B_{2},c^\B_{3}\},\mathcal{R}_\B)$ is 
the word algebra of binary strings. 
$\C=(\{c^\C_1,c^\C_2\},\mathcal{R}_\C)$,
where $\mathcal{R}_\C(c^\C_1)=2$ and $\mathcal{R}_\C(c^\C_2)=0$
is the free algebra of binary trees.
$\D=(\{c^\D_1,c^\D_2,c^\D_3\},\mathcal{R}_\D)$,
where $\mathcal{R}_\D(c^\D_1)=\mathcal{R}_\D(c^\D_2)=2$ and $\mathcal{R}_\D(c^\D_3)=0$
is the free algebra of binary trees with binary labels. Natural
numbers can be encoded by terms in $\U$:
$\inttou{0}= c_2^\U$ and $\inttou{n+1}=c_1^\U\inttou{n}$ for all $n$.
In the same vein, elements of $\{0,1\}^*$ are in one-to-one 
correspondence to terms in $\B$: $\bintob{\varepsilon}=c_3^\B$,
and for all $s\in\{0,1\}^*$, $\bintob{0 s}=c_1^\B\bintob{s}$ and
$\bintob{1s}=c_2^\B\bintob{s}$. When this does not cause ambiguity,
$\mathcal{C}_\A$ and $\mathcal{R}_\A$ will be denoted by
$\mathcal{C}$ and $\mathcal{R}$, respectively.

$\mathscr{A}$ will be a fixed, finite family $\{\A_1,\ldots,\A_n\}$ of free algebras 
whose constructor sets $\mathcal{C}_{\A_1},\ldots,\mathcal{C}_{\A_n}$
are assumed to be pairwise disjoint. We will hereby assume
$\U,\B,\C$ and $\D$ to be in $\mathscr{A}$. $\mathscr{K}_\mathscr{A}$ is the maximum
arity of constructors of free algebras in $\mathscr{A}$, i.e. the natural number
$$
\max_{\A\in\mathscr{A}}\max_{c\in\mathcal{C}_\A}\mathcal{R}_\A(c). 
$$
$\mathscr{E}_\A$ is the set of terms for the algebra $\A$, while
$\mathscr{E}_\mathscr{A}$ is the union of $\mathscr{E}_\A$ over all
algebras $\A$ in $\mathscr{A}$.

Programs will be written in a fairly standard lambda calculus with constants (corresponding
to free algebra constructors) and recursion. The latter will not be a combinator but
a term former, as in~\cite{Matthes03aml}. Moreover,
we will use a term former for conditional, keeping it distinct from the one for recursion.
This apparent redundancy is actually needed in presence of ramification (see,
for example,~\cite{Leivant93popl}). The language $\mathscr{M}_\mathscr{A}$ of \emph{terms} 
is defined by the following productions:
$$
M ::=
x                       \spb
c                       \spb
M M                    \spb
\lambda x.M             \spb
 M\;\llcu M,\ldots,M\rrcu\spb 
 M\;\llan M,\ldots,M\rran 
$$
where $c$ ranges over the constructors for the free algebras in $\mathscr{A}$.
Term formers $\cdot\;\llcu \cdot,\ldots,\cdot\rrcu$ and
$\cdot\;\llan \cdot,\ldots,\cdot\rran$ are \emph{conditional} and
\emph{recursion} term formers, respectively.\par
The language $\mathscr{T}_\mathscr{A}$ of \emph{types} is
defined by the following productions:
$$
A ::= \A^n\spb A\linear A
$$
where $n$ ranges over $\N$ and $\A$ ranges over $\mathscr{A}$.
Indexing base types is needed to define ramification conditions
as in~\cite{Leivant93popl}; $\A^n$, in
particular, is not a cartesian product.
The notation $A\linearo{n}B$ is defined by induction
on $n$ as follows: $A\linearo{0}B$ is just $B$, while
$A\linearo{n+1}B$ is $A\linear(A\linearo{n}B)$.
The {\it level} $V(A)\in\N$ of a type $A$
is defined by induction on $A$:
\begin{eqnarray*}
V(\A^n)&=&n;\\
V(A\linear B)&=&\max\{V(A),V(B)\}.
\end{eqnarray*}
When this does not cause ambiguity, we will denote
a base type $\A^n$ simply by $\A$.\par
The rules in Figure~\ref{figure:TYPEASSIGN}
define the assignment of types in $\mathscr{T}_\mathscr{A}$
to terms in $\mathscr{M}_\mathscr{A}$.
A type derivation $\pi$ with conclusion
$\Gamma\vdash M:A$ will be denoted by
$\pi:\Gamma\vdash M:A$. If there is
$\pi:\Gamma\vdash M:A$ then we will mark
$M$ as a \emph{typeable} term. 
A type derivation $\pi:\Gamma\vdash M:A$ is in \emph{standard form} if
the typing rule $W$ is used only when necessary, i.e. immediately
before an instance of $I_\linear$. We will hereby
assume to work with type derivations in standard form. This restriction
does not affect the class of typeable terms.\par
\begin{figure*}
\begin{center}
\input{typeassignment}
\caption{Type assignment rules}
\label{figure:TYPEASSIGN}
\end{center}
\end{figure*}
The {\it recursion depth} $R(\pi)$ of a
type derivation $\pi:\Gamma\vdash M:A$ is
the biggest number of $E^R_\A$ instances
on any path from the root to a leaf in $\pi$. 
The {\it highest tier} $I(\pi)$ of a type
derivation $\pi:\Gamma\vdash M:A$ is
the maximum integer $i$ such that there
is an instance
$$
\infer[E_\A^R]
{\Gamma,\Delta\vdash L\;\llan M_1,\ldots,M_n\rran:C}
{\pi_1 & \ldots & \pi_n & \Delta\vdash L:\A^i}
$$
of $E_\A^R$ inside $\pi$.\par
\emph{Values} are defined by the following productions:
\begin{eqnarray*}
V&::=&x\spb\lambda x.M\spb T;\\
T&::=&c\spb TT.
\end{eqnarray*}
where $c$ ranges constructors.
Reduction is weak and call-by-value. The reduction rule 
$\rightarrow$ on $\mathscr{M}_\mathscr{A}$ is given
in Figure~\ref{figure:NORMALIZATION}. 
\begin{figure*}
\begin{center}
\input{normalization}
\caption{Normalization on terms}
\label{figure:NORMALIZATION}
\end{center}
\end{figure*}
We will forbid firing a redex under an abstraction or inside a
recursion or a conditional. In other words, we will define
$\leadsto$ from $\rightarrow$ by the following set of rules:
$$
\begin{array}{ccc}
\infer{M\leadsto N}{M\rightarrow N}&
\infer{ML\leadsto NL}{M\leadsto N}&
\infer{LM\leadsto LN}{M\leadsto N}
\end{array}
$$
$$
\begin{array}{cc}
\infer{M\;\llcu L_1,\ldots,L_n\rrcu\leadsto N\;\llcu L_1,\ldots,L_n\rrcu}{M\leadsto N}&
\infer{M\;\llan L_1,\ldots,L_n\rran\leadsto N\;\llan L_1,\ldots,L_n\rran}{M\leadsto N}
\end{array}
$$ 
Redexes in the form $(\lambda x.M)V$ are called \emph{beta redexes}; those like
$t\;\llcu M_1,\ldots,M_{n}\rrcu$
are called \emph{conditional redexes}; those in the form
$t\;\llan M_1,\ldots,M_{n}\rran$
are \emph{recursive redexes}. The \emph{argument} of the
beta redex $(\lambda x.M)V$ is $V$, while that of
$t\;\llcu M_1,\ldots,M_{n}\rrcu$ and
$t\;\llan M_1,\ldots,M_{n}\rran$ is $t$.
As usual, $\leadsto^*$ and $\leadsto^+$ denote the
reflexive and transitive closure of $\leadsto$ and
the transitive closrue of $\leadsto$, respectively.

\begin{proposition}\label{prop:reducibility}
If $\vdash M:\A^n$, then the (unique) normal form of 
$M$ is a free algebra term $t$.
\end{proposition}
\begin{proof}
In this proof, terms from the grammar $T::=c\spb TT$ (where
$c$ ranges over constructors) are dubbed \emph{algebraic}. 
We prove the following stronger claim by
induction on $M$: if $\vdash M:A$ and $M$
is a normal form, then it must be a value.
We distinguish some cases:
\begin{varitemize}
  \item
  A variable cannot be typed in the empty context, so $M$ cannot
  be a variable.
  \item
  If $M$ is a constant or an abstraction, then it is a value by
  definition.
  \item
  If $M$ is an application $NL$, then there is a type $B$ such
  that both $\vdash N:B\linear A$ and $\vdash L:B$. By induction
  hypothesis both $N$ and $L$ must be values. But $N$ cannot be
  an abstraction (because otherwise $NL$ would be a redex) nor
  a variable (because a variable cannot be typed in the empty
  context). As a consequence, $N$ must be algebraic. Every
  algebraic term, however, has type $\A^i\linearo{n}\A^i$
  where $n\geq 0$. Clearly, this implies $n\geq 1$ and
  $B=\A^i$. This, in turn, implies $L$ to be algebraic (it
  cannot be a variable nor an abstraction). So, $M$ is
  itself algebraic.
  \item
  If $M$ is $N\;\llcu M_1,\ldots,M_{n}\rrcu$, then $N$ must
  be a value such that $\vdash N:\A^i$. As a consequence, it
  must be a free algebraic term $t$. But this is a contraddiction,
  since $M$ is assumed to be a value.
  \item
  If $M$ is $N\;\llan M_1,\ldots,M_{n}\rran$, then we 
  can proceed exactly as in the previous case.
\end{varitemize}
This concludes the proof, since the relation $\leadsto$ enjoys a one-step diamond
property (see~\cite{dallago06cie}).
\hfill $\Box$
\end{proof}
It should be now clear that
the usual recursion combinator $\mathbf{R}$ can be retrieved by putting
$\mathbf{R}=\lambda x.\lambda y_1.\ldots.\lambda y_n.x\;\llan y_1,\ldots,y_n\rran$.

The \emph{size} $|M|$ of a term $M$ is defined as follows
by induction on the structure of $M$:
\begin{eqnarray*}
|x|=|c|&=&1\\
|\lambda x.M|&=&|M|+1\\
|MN|&=&|M|+|N|\\
|M\llan M_1,\ldots,M_n\rran|=|M\llcu M_1,\ldots,M_n\rrcu|&=&|M|+|M_1|+\ldots+|M_n|+n
\end{eqnarray*}
Notice that, in particular, $|t|$ equals the number of constructors in $t$
for every free algebra term $t$.

\section{Subsystems}\label{sect:subsystems}
The system, as it has been just defined, is equivalent
to G\"odel System $\mathsf{T}$ and, as a consequence, its expressive power
equals the one of first-order arithmetic. We are here 
interested in two different conditions on
programs, which can both be expressed as
constraints on the underlying type-system:
\begin{varitemize}
  \item
  First of all, we can selectively enforce
  \emph{linearity} by limiting the applicability of 
  contraction rule $C$ to types in a class
  $\mathsf{D}\subseteq\mathscr{T}_\mathscr{A}$.
  Accordingly, the constraint
  $\mathit{cod}(\Gamma_i)\subseteq\mathsf{D}$ 
  must be satisfied in rule $E^R_\A$ (for
  every $i\in\{1,\ldots,n\}$). 
  In this way, we obtain a system $\h{\mathsf{D}}$.
  As an example, $\h{\emptyset}$ is 
  a system where rule $C$ is not allowed on any type and
  contexts $\Gamma_i$ are always empty in
  rule $E^R_\A$.
  \item 
  Secondly, we can introduce a \emph{ramification}
  condition on the system. This can be done in a
  straightforward way by adding the premise
  $m>V(C)$ to rule $E^R_\A$. This corresponds to
  impose the tier of the recurrence argument to
  be strictly higher than the tier of the result
  (analogously to Leivant~\cite{Leivant93popl}).
  Indeed, $m$ is the integer indexing the type of the
  recurrence argument, while $V(C)$ is the maximum
  integer appearing as an index in $C$, which is the
  type of the result. For every system $\h{\mathsf{D}}$, 
  we obtain in this way a ramified system $\rh{\mathsf{D}}$.
\end{varitemize}
The constraint $\mathit{cod}(\Gamma_i)\subseteq\mathsf{D}$
in instances of rule $E^R_\A$ is needed to preserve linearity
during reduction: if $c_i^\A t\;\llan M_1,\ldots,M_{k(\A)}\rran$ is 
a recursive redex where $M_i$ has a free variable $x$ of
type $A\notin\mathsf{D}$, firing the redex would produce a term with
two occurrences of $x$.

Let us define the following two classes of types:
\begin{eqnarray*}
\mathsf{W}&=&\{\A^n\spb\mbox{\emph{$\A\in\mathscr{A}$ is a word algebra}}\},\\
\mathsf{A}&=&\{\A^n\spb \A\in\mathscr{A}\}.
\end{eqnarray*}
In the rest of this paper, we will investigate the expressive
power of some subsytems $\mathsf{H(D)}$ and $\mathsf{RH(D)}$
where $\mathsf{D}\subseteq\mathsf{A}$. The following table reports the obtained results:
\begin{center}
\begin{tabular}{|c|c|c|c|}
\cline{2-4}
\multicolumn{1}{c|}{} & $\mathsf{A}$ & $\mathsf{W}$ & $\emptyset$ \\ \hline\hline
$\h{\cdot}$ & $\mathbf{FR}$ & $\mathbf{FR}$ & $\mathbf{FR}$ \\ \hline
$\rh{\cdot}$ & $\mathbf{FE}$ & $\mathbf{FP}$ & $\mathbf{FP}$ \\
\hline\hline
\end{tabular}
\end{center}
Here, $\mathbf{FP}$ (respectivel{}y, $\mathbf{FE}$)  is the class of functions which can 
be computed in polynomial (respectively, elementary) time. $\mathbf{FR}$, on the
other hand, is the class of (first-order) primitive recursive functions, which equals the class of 
functions which can be computed in time bounded by a primitive recursive function. 
For example, $\rh{\mathsf{A}}$ is proved sound and complete with respect to elementary
time, while $\h{\emptyset}$ is shown to capture (first-order) primitive recursion.

Forbidding contraction on higher-order types is quite common and has been extensively
used as a tool to restrict the class of representable 
functions inside System $\mathsf{T}$~\cite{hofmann00safe,Bellantoni00apal,
Leivant99csl}. The correspondence between $\rh{\mathsf{W}}$ and $\mathbf{FP}$ is well known from the 
literature~\cite{hofmann00safe,Bellantoni00apal}, although in a slightly
different form. To the author's knowledge, all the other charaterization results are 
novel. Similar results can be ascribed
to Leivant and Marion~\cite{leivant94csl,Leivant99apal}, but they do not take 
linearity constraints into account.

Notice that, in presence of ramification, going from $\mathsf{W}$ to $\mathsf{A}$
dramatically increases the expressive power, while going from $\mathsf{W}$
to $\emptyset$ does not cause any loss of expressivity. The ``phase-transition''
occurring when switching from $\rh{\mathsf{W}}$ to $\rh{\mathsf{A}}$ is really
surprising, since the only difference between these two systems are the class
of types to which linearity applies: in one case we only have word algebras, while 
in the other case we have all free algebras.

\section{Algebraic Context Semantics}\label{sect:acs}
In this section, we will introduce algebraic context semantics,
showing how bounds on the normalization time of a any term
$M$ can be inferred from its semantics. 

The first result we need relates the complexity of normalizing any given term $M$ to the
size of free algebra terms appearing as subterms of reducts
of $M$. The \emph{algebraic potential size} $A(M)$ of a typable term
$M$ is the maximum natural number $n$ such that
$M\leadsto^* N$ and there is a redex in
$N$ whose argument is a free algebra term $t$ with 
$|t|=n$. Since the calculus is strongly 
normalizing, there is always a finite bound
to the size of reducts of a term and, as a
consequence, the above definition is well-posed.
According to the following result,
the algebraic potential size of a term $M$
such that $\pi:\Gamma\vdash_{\h{\mathsf{A}}} M:A$ is
an overestimate on the time needed to normalize
the term (modulo some polynomials that only
depends on $R(\pi)$): 
\begin{proposition}\label{prop:justification}
For every $d\in\N$ there are polynomials $p_d,q_d:\N^2\rightarrow\N$
such that whenever $\pi:\Gamma\vdash_{\h{\mathsf{A}}} M:A$ and
$M\leadsto^n N$, then $n\leq p_{R(\pi)}(|M|,A(M))$
and $|N|\leq q_{R(\pi)}(|M|,A(M))$.
\end{proposition}
\begin{proof}
Let us first observe that the number of recursive
redexes fired during normalization of $M$ is bounded
by $s_{R(\pi)}(|M|,A(M))$, where
$$
s_{d}(x,y)=xy^d
$$
Indeed, consider subterms of $M$ in the form $L \llan N_1,\ldots,N_k\rran$. 
Clearly, there are at most $|M|$ such terms. Moreover, each such
subterm can result in at most $|A(M)|^{R(\pi)}$ recursive redexes.
Indeed, it can be copied at most $|A(M)|^{R(\pi)-1}$
times, and each copy can itself result in $|A(M)|$ recursive redexes. 
Now, notice that firing a beta or a conditional 
redex does not increase the number of variable occurrences
in the term. Conversely, firing
a recursive redex can make it bigger by at most
$|M|$. We can conclude that the number of
beta redexes in the form $(\lambda x.M)t$
(let us call them \emph{algebraic redexes})
is at most $|M|s_{R(\pi)}(|M|,A(M))$ and, moreover,
they can make the term to increase in size
by at most $A(M)|M|s_{R(\pi)}(|M|,A(M))$ altogether.
Firing a recursive redex
$$
c_i(t_1,\ldots,t_{\mathcal{R}(c_i)})
\llan M_{c_1},\ldots,M_{c_k}\rran
$$
can make the size of the underlying term to 
increase by $r_{R(\pi)}(|M|,A(M))$ where
$$
r_{R(\pi)}(x,y)=\mathscr{K}_\mathscr{A}(y+x+xy).
$$ 
Indeed:
\begin{eqnarray*}
&&|M_{c_i}\;t_1\cdots t_{\mathcal{R}(c_i)}
  (t_1\;\llan M_{c_1},\ldots,M_{c_k}\rran)
  \cdots
  (t_{\mathcal{R}(c_i)}\;\llan M_{c_1},\ldots,M_{c_k}\rran)|\\
&=&|M_{c_i}\;t_1\cdots t_{\mathcal{R}(c_i)}|+|t_1\;\llan M_{c_1},\ldots,M_{c_k}\rran)|+\ldots+
  |t_{\mathcal{R}(c_i)}\;\llan M_{c_1},\ldots,M_{c_k}\rran|\\
&\leq&|c_i(t_1,\ldots,t_{\mathcal{R}(c_i)})\llan M_{c_1},\ldots,M_{c_k}\rran|+
  \sum_{i=1}^{\mathcal{R}(c_i)}(|t_i|+|M_{c_1}|+\ldots+|M_{c_k}|+k)\\
&\leq&|c_i(t_1,\ldots,t_{\mathcal{R}(c_i)})\llan M_{c_1},\ldots,M_{c_k}\rran|+
  \mathscr{K}_\mathscr{A}(A(M)+|M|+A(M)|M|)
\end{eqnarray*}
because $|M_{c_1}|+\ldots+|M_{c_k}|+k$ is bounded by
$|M|+A(M)|M|$ and $|c_i(t_1,\ldots,t_{\mathcal{R}(c_i)})|$
is bounded by $A(M)$.
We can now observe that firing any redex other than algebraic
or recursive ones makes the size of the term to strictly decrease. As
a consequence, we can argue that
\begin{eqnarray*}
q_d(x,y)&=&x+s_d(x,y)xy+s_d(x,y)r_d(x,y);\\
p_d(x,y)&=&s_d(x,y)x+s_d(x,y)+q_d(x,y).
\end{eqnarray*}
This concludes the proof.\qed
\end{proof}
Observe that in the statement of Proposition~\ref{prop:justification}, it is 
crucial to require $M$ to be typable in $\h{\mathsf{A}}$. Indeed, it is quite
easy to build simply-typed (pure) lambda terms which have exponentially big
normal forms, although having null algebraic potential size.

In the rest of this section, we will develop a semantics, derived from
context semantics~\cite{Gonthier92popl} and dubbed \emph{algebraic context semantics}. 
We will then use it to give bounds to the algebraic potential
size of terms in subsystems we are interested in and
use Proposition~\ref{prop:justification} to derive time bounds.

Consider the term 
$$
\mathbf{UnAdd}\equiv \lambda x.\lambda y.x\llan \lambda w.\lambda z.c_1^\U z,y\rran.
$$
Clearly, $\mathbf{UnAdd}\inttou{n}\inttou{m}\leadsto^*\inttou{n+m}$.
$\mathbf{UnAdd}\inttou{1}\inttou{1}$ will be used as a reference example throughout this section.
A type derivation $\sigma$ for $\mathbf{UnAdd}\inttou{1}\inttou{1}$ is the following one:
$$
\infer
{\vdash \mathbf{UnAdd}\inttou{1}\inttou{1}:\U^0}
{
  \infer
  {\vdash \mathbf{UnAdd}\inttou{1}:\U^0\linear\U^0}
  {
    \infer
    {\mathbf{UnAdd}:\U^1\linear\U^0\linear\U^0}
    {
      \infer{x:\U^1\vdash\lambda y.x\llan \lambda w.\lambda z.c_1^\U z,y\rran:\U^0\linear\U^0}
      {
        \infer{x:\U^1,y:\U^0\vdash x\llan \lambda w.\lambda z.c_1^\U z,y\rran:\U^0}
        {
          \infer{\vdash\lambda w.\lambda z.c_1^\U z:\U^1\linear\U^0\linear\U^0}
          {
            \infer{w:\U^1\vdash\lambda z.c_1^\U z:\U^0\linear\U^0}
            { 
              \infer{w:\U^1,z:\U^0\vdash c_1^\U z:\U^0}
              {
                \infer{z:\U^0\vdash c_1^\U z:\U^0}
                {
                  \infer{\vdash c_1^\U:\U^0\linear\U^0}{}
                  &
                  \infer{z:\U^0\vdash z:\U^0}{}
                }
              }
            }
          }
          &
          \infer{y:\U^0\vdash y:\U^0}{}
          &
          \infer{x:\U^1\vdash x:\U^1}{}
        }
      }
    }
    &
    \eta_1:\vdash \inttou{1}:\U^1
  }
  &
  \eta_0\vdash \inttou{1}:\U^0
}
$$
where $\eta_0$ and $\eta_1$ are defined in the obvious way.

We will study the context semantics of interaction graphs, which are graphs 
corresponding to type derivations. Notice that we will not use
interaction graphs as a virtual machine computing normal forms 
--- they are merely a tool facilitating the study of language dynamics.
More precisely, we will put every type derivation $\pi$ in correspondence
to an interaction graph $G_\pi$. The context semantics of $G_\pi$
will be a set of trees $T(G_\pi)$ such that every tree $T$ in $T(G_\pi)$
can be associated to a term $t=L(T)\in\mathscr{E}_\mathscr{A}$. 
If $\pi:\Gamma\vdash M:A$, then $T(G_\pi)$ keeps track of the
normalization of $M$ in the following sense: for every $t$ 
appearing as an argument of a reduct of $M$, there is a 
tree $T\in T(G_\pi)$ such that $t=L(T)$. Proving this property, called
completeness, is the aim of Section~\ref{subsect:completeness}. Completeness,
together with Proposition~\ref{prop:justification}, is exploited
in Section~\ref{sect:cn}, where bounds on normalization time for
classes of terms are inferred.

\begin{figure}
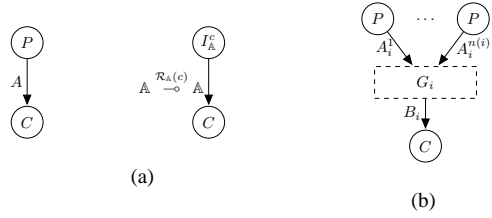

\begin{center}
\subfigure[]{
  \begin{minipage}[c]{13.2pt}
    \centering\scalebox{0.6}{\epsfbox{figure.8}}
  \end{minipage}
  \hspace{10pt}
  \begin{minipage}[c]{74.4pt}
    \centering\scalebox{0.6}{\epsfbox{figure.9}}
  \end{minipage}
}
\hspace{20pt}
\subfigure[]{
  \begin{minipage}[c]{49.8pt}
    \centering\scalebox{0.6}{\epsfbox{figure.2}}
  \end{minipage}
}
\caption{Base cases.}
\label{fig:baseinduction}
\end{center}
\end{figure}

\begin{figure*}
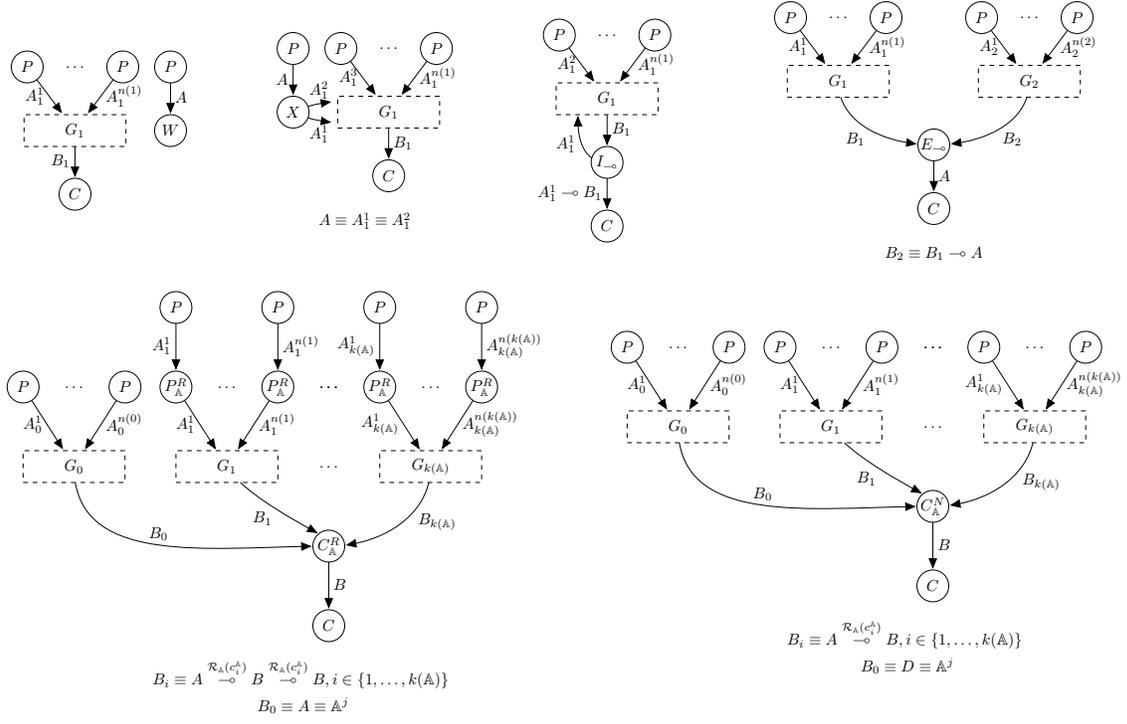
 
\begin{center}
  \begin{minipage}[c]{73.2pt}
    \centering\scalebox{0.6}{\epsfbox{figure.10}}
  \end{minipage} \hspace{20pt}
  \begin{minipage}[c]{79.2pt}
    \centering\scalebox{0.6}{\epsfbox{figure.25}}
  \end{minipage} \hspace{20pt}
  \begin{minipage}[c]{52.2pt}
    \centering\scalebox{0.6}{\epsfbox{figure.3}}
  \end{minipage} \hspace{20pt}
  \begin{minipage}[c]{146.8pt}
    \centering\scalebox{0.6}{\epsfbox{figure.4}}
  \end{minipage}\\\vspace{10pt}
  \begin{minipage}[c]{202.2pt}
    \centering\scalebox{0.6}{\epsfbox{figure.7}}
  \end{minipage} \hspace{20pt}
  \begin{minipage}[c]{196.5pt}
    \centering\scalebox{0.6}{\epsfbox{figure.13}}
  \end{minipage}
\end{center}
\caption{Inductive cases}
\label{fig:induction}
\end{figure*}

Let $\mathscr{L}_\mathscr{A}$ be the set 
$$
\{W,X,I_\linear,E_\linear,P,C\}\cup\bigcup_{\A\in\mathscr{A}}\{C^N_\A,P^R_\A,C^R_\A\}
\cup\bigcup_{\A\in\mathscr{A}}\bigcup_{c\in\mathcal{C}_\A}\{I^c_\A\}.
$$ 
An \emph{interaction graph} is a graph-like structure $G$. It
can be defined inductively as follows: an interaction graph is either the
graph in Figure~\ref{fig:baseinduction}(a) or one of those in Figure~\ref{fig:induction} where 
$G_0,G_1,\ldots,G_{k(\A)}$ are themselves proof-nets as in Figure~\ref{fig:baseinduction}(b).
If $G$ is an interaction graph, then $V_G$ denotes the set of vertices
of $G$, $E_G$ denotes the set of directed edges of $G$,
$\alpha_G$ is a labelling function mapping every vertex in $V_G$ to an 
element of $\mathscr{L}_\mathscr{A}$ and $\beta_G$ maps every edge in $E_G$ to a 
type in $\mathscr{T}_\mathscr{A}$.
$\mathscr{G}_\mathscr{A}$ is the set of all interaction graphs. 

\begin{figure*}
\begin{center}
\scalebox{0.6}{\epsfbox{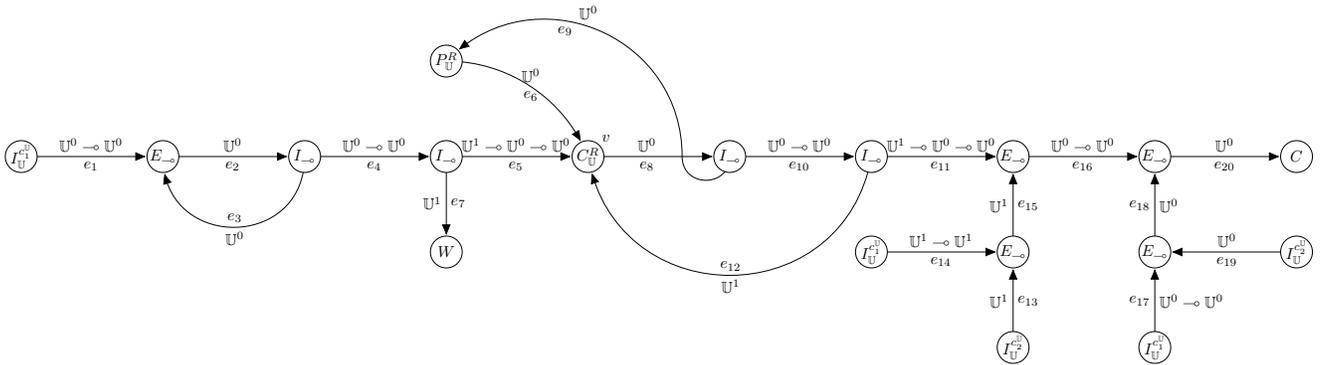}}
\caption{The interaction graph corresponding to a type derivation for $\mathbf{UnAdd}\inttou{1}\inttou{1}$}
\label{fig:unadd}
\end{center}
\end{figure*}

Notice that each of the rules in figures~\ref{fig:baseinduction}(a) and~\ref{fig:induction}
closely corresponds to a typing rule. Given a type derivation $\pi$, we can then build an 
interaction graph $G_\pi$ corresponding to $\pi$. For example, Figure~\ref{fig:unadd}
reports an interaction graph $G_\sigma$ where
$\sigma:\;\vdash \mathbf{UnAdd}\inttou{1}\inttou{1}:\U^0$. Let us observe that
 if $\pi:\Gamma\vdash M:A$ is in standard form, then the size $|G_\pi|$ of $G_\pi$ is proportional to $|M|$. 
\par
Nodes labelled with $C$ (respectively, $P$) mark
the conclusion (respectively, the premises) of
the interaction graph. Notice that the rule corresponding to
recursion (see Figure~\ref{fig:induction}) allows 
seeing interaction graphs
as nested structures, where nodes labelled
with $C^R_\A$ and $P^R_\A$ delimit a \emph{box}, similarly to
what happens in linear logic proof-nets. If $e\in E_G$,
then the \emph{box-premise} of $e$, denoted
$\theta_G(e)$, is the vertex labelled with $C^R_\A$ 
delimiting the box in which $e$ is contained 
(if such a box exists, otherwise $\theta_G(e)$ is undefined). 
If $v\in V_G$, the box-premise of $v$, denoted
$\theta_G(v)$ is defined similarly.
In our example (see Figure~\ref{fig:unadd}), 
$\theta_G(e_i)$ equals $v$ for every $i\leq 7$ and is 
otherwise undefined.
If $v$ is a vertex with $\alpha_G(v)=C^R_\A$, then
the \emph{recursive premise} of $v$, denoted 
$\rho_G(v)$, is the edge incident to $v$ 
and coming from outside the box. 
In our example, $\rho_G(v)$ is $e_{12}$.

Defining algebraic context semantics requires a number
of auxiliary concepts, like the one of a term context
and the one of a type context. The set $\mathscr{N}_\A$ 
of \emph{term contexts} for $\A$ is defined as follows:
\begin{varitemize}
  \item
  $[\cdot]\in\mathscr{N}_\A$
  \item
  If $t_1,\ldots,t_m\in\mathscr{E}_\A$,
  $u\in\mathscr{N}_\A$ and  
  $c$ is a constructor of $\A$ with arity $m+1$, then
  $c t_1\ldots t_{i}ut_{i+1}\ldots t_m\in\mathscr{N}_\A$,
  for every $i\in\{0,m\}$.
\end{varitemize}
In other words, elements of $\mathscr{N}_\A$
are terms with a hole inside them. 
If $u\in\mathscr{N}_\A$ and 
$t\in\mathscr{E}_\A$, then
$u[t]\in\mathscr{E}_\A$ is obtained by replacing
the hole inside $u$ with $t$ in the
obvious way. Similarly, if $u,t\in\mathscr{N}_\A$,
then $u[t]$ will be a context in $\mathscr{N}_\A$.
$\mathscr{N}_\mathscr{A}$ is defined
in the usual way.\par
The classes $\mathscr{C}_\mathscr{A}^+$ and
$\mathscr{C}_\mathscr{A}^-$ of positive and negative 
\emph{type contexts} are defined as follows:
\begin{varitemize}
  \item
  $[\cdot]\in\mathscr{C}_\mathscr{A}^+$ is a positive type context.  
  \item
  If $L\in\mathscr{C}_\mathscr{A}^+$ 
  and $A$ is a type, then $L\linear A\in\mathscr{C}_\mathscr{A}^-$
  and $A\linear L\in\mathscr{C}_\mathscr{A}^+$.
  \item
  If $L\in\mathscr{C}_\mathscr{A}^-$ 
  and $A$ is a type, then $L\linear A\in\mathscr{C}_\mathscr{A}^+$
  and $A\linear L\in\mathscr{C}_\mathscr{A}^-$.
\end{varitemize}
$\mathscr{C}_\mathscr{A}$ is just $\mathscr{C}^+_\mathscr{A}\cup\mathscr{C}^-_\mathscr{A}$.
A context $L$ is a \emph{focus} for $A$ if there are
a free algebra and a natural number $i$ such that $\A$ and$A\equiv L[\A^i]$.

Given an interaction graph $G$, its context semantics
is given by a set $T(G)$ of trees. Vertices
of trees in $T(G)$ are labelled with contexts in $S(G)$,
where
\begin{eqnarray*}
  S(G)&=&\mathscr{E}_\mathscr{A}\times E_G\times C(G)\times\mathscr{C}_\mathscr{A};\\
  C(G)&=&(\mathscr{N}_\mathscr{A}\times\mathscr{E}_\mathscr{A}\times V_G)^*.
\end{eqnarray*}
In other words, elements of $T(G)$ are couples $(l,f)$ where $l\in S(G)$,
$f:\{1,\ldots,k\}\rightarrow T(G)$ and $k\in\N$ Elements of $C(G)$ are
dubbed \emph{stacks}. To any tree $T\in T(G)$ we can put in correspondence 
a term $t=L(T)\in\mathscr{E}_\mathscr{A}$ by picking up the first 
component of the tuple labelling its root.

We will define $T(G)$ by closure conditions. Most of them will
be like the following: if $(l_1,f_1),\ldots,(l_k,f_k)\in T(G)$,
then $(l,f)\in T(G)$, where $\forall i\in\{1,\ldots,k\}.f(i)=(l_i,f_i)$; 
These conditions are denoted by writing $l\leftarrow (l_1,f_1),\ldots,(l_k,f_k)$
or $l\leftarrow l_1,\ldots,l_k$ if this does not cause ambiguity.
More specifically, any closure condition we are going to define will be
in one of the following two forms:
\begin{varitemize}
  \item
    In the form $(t,e,U,L)\leftarrow(t,g,V,M)$. In other words, the root
    of the newly defined tree will have just one immediate descendant and
    the term labelling the first component of the root will be the
    same as the term labelling the first component of its immediate descendant.
  \item
    In the form $(ct_1\ldots t_n,e,U,L)\leftarrow(t_1,g_1,V_1,M_1),\ldots
    (t_n,g_n,V_n,M_n)$. In this case, the root of the newly defined tree
    will have $n$ immediate descendants and the term labelling the first
    component of the root will be built by applying a constructor to terms
    appearing as first components of its immediate descendants.
\end{varitemize}
As a consequence, if $t=u[s]$, then there is at least one subtree $S$ of $T$
such that $L(S)=s$ and $S$ somehow corresponds to $u$. 
We will denote the smallest of such subtrees by $B(T,u,s)$. Observe
the root of $B(T,u,s)$ is the last node we find when travelling 
from the root of $T$ toward its leaves and being guided by $u$.

\begin{figure}
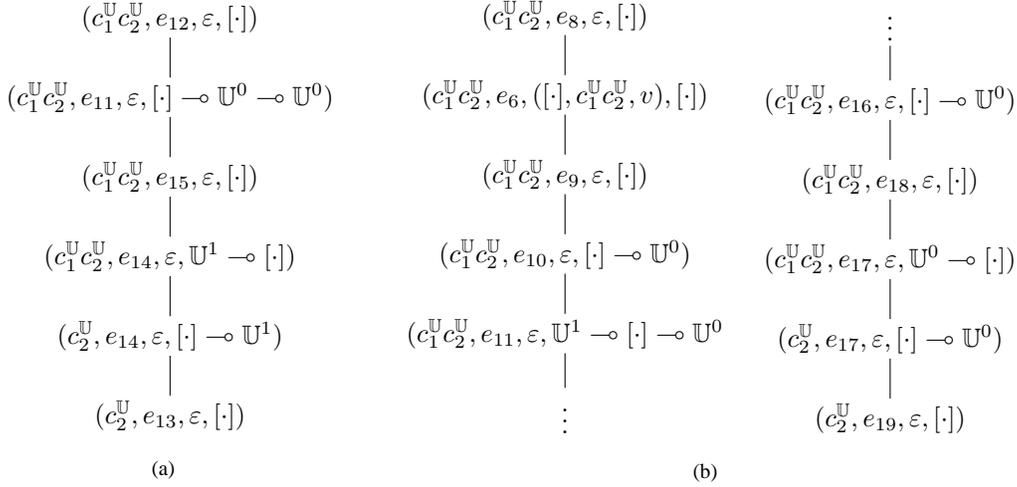

\begin{center}
\subfigure[]{
  \begin{minipage}[c]{124pt}
    \centering\scalebox{1}{\epsfbox{figure.47}}
  \end{minipage}}
  \hspace{20pt}
\subfigure[]{
  \begin{minipage}[c]{120pt}
    \centering\scalebox{1}{\epsfbox{figure.48}}
  \end{minipage}
  \hspace{10pt}
  \begin{minipage}[c]{96pt}
    \centering\scalebox{1}{\epsfbox{figure.49}}
  \end{minipage}}
  \caption{Examples of trees.}
\label{fig:treeexamples}
\end{center}
\end{figure}

Formally, $T(G)$ is defined as the smallest set
satisfying two families of closure conditions: 
\begin{varitemize}
  \item 
  Vertices of $G$ with labels
  $I_\linear$ ,$E_\linear$, $X$ $I^c_\A$ and $C^N_\A$ induce
  closure conditions on $T(G)$. These conditions are detailed 
  in Table~\ref{table:sgone}.
  \item
  Every vertex with labels $C^R_\A$ and  $P^R_\A$ forces $T(G)$ 
  to satisfy more complex closure conditions, 
  as reported in Table~\ref{table:sgtwo}. 
\end{varitemize}
\begin{table*}
\begin{center}
\caption{Closure conditions}\label{table:sgone}
\vspace{10pt}
\begin{tabular}{|c|c|}\hline\hline
\begin{minipage}[c]{2.5cm}
\vspace{.3cm}
\centering\scalebox{0.60}{\epsfbox{figure.26}}\\
\vspace{.3cm}
\end{minipage}
 &
\begin{minipage}[c]{9cm}
\begin{eqnarray*}
(t,e,U,P_A)&\leftarrow&(t,h,U,P_A\linear B)\\
(t,h,U,N_A\linear B)&\leftarrow&(t,e,U,N_A)\\
(t,h,U,A\linear P_B)&\leftarrow&(t,g,U,P_B)\\
(t,g,U,N_B)&\leftarrow&(t,h,U,A\linear N_B)\\
\end{eqnarray*}
\end{minipage}
\\ \hline
\begin{minipage}[c]{2.5cm}
\vspace{.3cm}
\centering\scalebox{0.60}{\epsfbox{figure.27}}\\
\vspace{.3cm}
\end{minipage}
 &
\begin{minipage}[c]{9cm}
\begin{eqnarray*}
(t,e,U,P_A\linear B)&\leftarrow&(t,g,U,P_A)\\
(t,g,U,N_A)&\leftarrow&(t,e,U,N_A\linear B)\\
(t,e,U,A\linear N_B)&\leftarrow&(t,h,U,N_B)\\
(t,h,U,P_B)&\leftarrow&(t,e,U,A\linear P_B)\\
\end{eqnarray*}
\end{minipage}
\\ \hline
\begin{minipage}[c]{2.5cm}
\vspace{.3cm}
\centering\scalebox{0.60}{\epsfbox{figure.30}}\\
\vspace{.3cm}
\end{minipage}
 &
\begin{minipage}[c]{9cm}
\begin{eqnarray*}
(t,e,U,[\cdot])&\leftarrow&(t,h,U,[\cdot])\\
(t,g,U,[\cdot])&\leftarrow&(t,h,U,[\cdot])\\
\end{eqnarray*}
\end{minipage}
\\ \hline
\begin{minipage}[c]{2.5cm}
\vspace{1.2cm}
\centering\scalebox{0.60}{\epsfbox{figure.35}}\\
\vspace{1.2cm}
\end{minipage}
 &
\begin{minipage}[c]{9cm}
  Let
  $T=((u[t],\rho_G(v),U,[\cdot]),f)\in T(G)$
  and $\theta_G(e)=v$. Then
  $$
  (c,e,(u,t,v)U,[\cdot])\leftarrow\\
  $$
  If $\theta_G(e)$ is undefined, then
  $$
  (c,e,\varepsilon,[\cdot])\leftarrow\\
  $$
\end{minipage}
\\ \hline
\begin{minipage}[c]{2.5cm}
\vspace{.9cm}
\centering\scalebox{0.60}{\epsfbox{figure.31}}\\
\vspace{.9cm}
\end{minipage}
 &
\begin{minipage}[c]{9cm}
$$
\begin{array}{c}
(c t_1\ldots t_{\ca{\A}{c}},e,U,\A\linearo{n}[\cdot])\leftarrow\\
(t_1,e,U,\A\linearo{0}[\cdot]\linear(\A\linearo{n-1}\A)),\\
\vdots \\
(t_{\ca{\A}{c}},e,U,\A\linearo{n-1}[\cdot]\linear(\A\linearo{0}\A))
\end{array}
$$
\end{minipage}
\\ \hline
\begin{minipage}[c]{2.5cm}
\vspace{.3cm}
\centering\scalebox{0.60}{\epsfbox{figure.33}}\\
\vspace{.3cm}
\end{minipage}
 &
\begin{minipage}[l]{9cm}
Let $T=((c_it_1\ldots t_{n_i},e_0,U,[\cdot]),f)\in T(G)$ and
$S_j=$ $B(T,c_it_1\ldots t_{j-1}[\cdot]t_{j+1}\ldots t_{n_i},t_j)$. Then
\begin{eqnarray*}
  (t,g,U,P_A)&\leftarrow&(t,e_i,U,\A\linearo{n_i}P_A)\\
  (t,e_i,U,\A\linearo{n_i}N_A)&\leftarrow&(t,g,U,N_A)\\
  (t_i,e_j,U,\A\linearo{n_i-j}[\cdot]\linear\A\linearo{j-1}A)&\leftarrow& S_j
\end{eqnarray*}
\end{minipage}
\\ \hline\hline
\end{tabular}
\end{center}
\end{table*}
\begin{table*}
\caption{Closure conditions}\label{table:sgtwo}
\vspace{10pt}
\begin{center}
\begin{tabular}{|c|c|}\hline\hline
\begin{minipage}[c]{2.5cm}
\vspace{2.5cm}
\centering\scalebox{0.60}{\epsfbox{figure.32}}\\
\vspace{2.5cm}
\end{minipage}
 &
\begin{minipage}[l]{12cm}
Let $T=((t,e_0,U,[\cdot]),f)\in T(G)$. If $t=u[s]$ and 
$s=c_i t_1\ldots t_{k-1} (c_j s_1\ldots s_{n_j})t_{k+1}\ldots t_{n_i}$, then
  $$
  \begin{array}{c}
    (r,e_i,(u,s,v)U,\A\linearo{n_i}(A\linearo{n_i-k}P_A
      \linear(A\linearo{k-1}A)))\leftarrow\\
    (r,e_j,(u[c_i t_1\ldots t_{k-1}[\cdot] t_{k+1}\ldots t_{n_i}],t_k,v)U,
        \A\linearo{n_j}(A\linearo{n_j}P_A))\\
    (r,e_j,(u[c_i t_1\ldots t_{k-1}[\cdot] t_{k+1}\ldots t_{n_i}],t_k,v)U,
        \A\linearo{n_j}(A\linearo{n_j}N_A))\leftarrow\\
    (r,e_i,(u,s,v)U,\A\linearo{n_i}(A\linearo{n_i-k}N_A
      \linear(A\linearo{k-1}A)))
  \end{array}
  $$
If $t=c_i t_1,\ldots,t_{n_i}$, then
  $$
  \begin{array}{c}
    (r,g,U,P_A)\leftarrow(r,e_i,([\cdot],t,v)U,\A\linearo{n_i}(A\linearo{n_i}P_A))\\
    (r,e_i,([\cdot],t,v)U,\A\linearo{n_i}A\linearo{n_i}N_A)\leftarrow(r,g,U,N_A)
  \end{array}
  $$
If $t=u[s]$, $s=c_it_1\ldots t_{n_i}$ and $S_j=B(T,u[c_it_1,\ldots,t_{j-1}[\cdot]t_{j+1}\ldots t_{n_i}],t_j)$, then
  $$
  (t_j,e_i,(u,s,v)U,\A\linearo{j-1}[\cdot]\linear \A\linearo{n_i-j}(A\linearo{n_i} A))
  \leftarrow S_j
  $$
\end{minipage}
\\ \hline
\begin{minipage}[c]{2.5cm}
\vspace{.3cm}
\centering\scalebox{0.60}{\epsfbox{figure.34}}\\
\vspace{.3cm}
\end{minipage}
 &
\begin{minipage}[l]{12cm}
Let
$T=((u[s],\rho_G(v),U,[\cdot]),f)\in T(G)$
and $\theta_G(g)=v$. Then
$$
(t,g,(u,s,v)U,[\cdot])\leftarrow(t,e,U,[\cdot])
$$
\end{minipage}
\\ \hline\hline
\end{tabular}
\end{center}
\end{table*}
In tables~\ref{table:sgone} and~\ref{table:sgtwo}, $P_A$ (respectively, $N_A$)
ranges over positive (respectively, negative) focuses for $A$, while
$P_B$ (respectively, $N_B$) ranges over positive (respectively, negative) focuses for $B$.
In Figure~\ref{fig:treeexamples}, we report two trees in $T(G_\sigma)$, where
$\sigma:\mathbf{UnAdd}\inttou{1}\inttou{1}:\U^0$.

Branches of trees in $T(G)$ correspond to paths inside $G$, i.e.
finite sequences of consecutive edges of $G$. The path corresponding to a branch
in $G$ can be retrieved by considering the second component
of tuples labelling vertices in the branch. The third 
and fourth components serve as contexts and are necessary to
build the tree in a correct way. Indeed, this way of building
trees by traversing paths is reminiscent of token machines 
in the context of game semantics and geometry of
interaction (see~\cite{danos96lics,Danos99tcs}). 
Using this terminology, we can informally describe the
components of a tuple as follows:
\begin{varitemize}
  \item
  The first component is a value carried by the token; it is modified
  when crossing a node labelled with $I^c_\A$;
  \item
  the third one is a stack and
  can only be changed by traversing a node labelled with 
  $C^R_\A$ or $P^R_\A$;
  \item
  the fourth component is a type context guiding the
  travel of the token. As we are going to show in the 
  following, the fourth component is always a focus for the type 
  labelling the current edge (which can be found
  in the second component).
\end{varitemize}
Some observations about the closure conditions in
tables~\ref{table:sgone} and~\ref{table:sgtwo} are 
now in order:
\begin{varitemize}
\item
  The only way of proving a one-node tree to be 
  in $T(G)$ consists in applying 
  the closure condition induced by a vertex $w$ labelled 
  with $I^c_\A$, where $\mathcal{R}_\A(c)=0$. Notice that,
  if $\theta_G(w)$ is defined (i.e. $w$ is inside a box), 
  we must check the existence of another (potentially big)
  tree $T$. Similarly when we want to ``enter'' a box
  by traversing a vertex $w$ labelled with $P^R_\A$.
\item
  Closure conditions induced by vertices labelled with
  $C^R_\A$ are quite complicated. Consider one such vertex
  $w$. First of all, a preliminary
  condition to be checked is the existence of a node $T$
  such that the second component of the tuple labelling
  the root of $T$ is the recursive premise of $w$. 
  The existence of $T$ certifies that exactly $|L(T)|$
  copies of the box under consideration will be produced
  during reduction, each of them corresponding to a 
  tuple $(u,s,w)$ where $u[s]=L(T)$. The vertex $w$ induce
  five distinct closure rules. The first
  two rules correspond to paths that come from the interior
  of the box under consideration and stay inside the same
  box: we go from one copy of the box to another one and,
  accordingly, the leftmost element of the underlying
  stack is changed. The third and fourth rules correspond
  to paths that enter or exit the box from its conclusion:
  an element of $C(G)$ is either popped from the underlying
  stack (when exiting the box) or pushed into it (when
  enterint the box). The last rule is definitely the
  trickiest one. First of all, remember that $L(T)$ 
  represents an argument to the recursion corresponding to 
  $w$. If we look at the reduction rule for recursive redexes, we 
  immediately realize that subterms of this argument should
  be passed to the bodies of the recursion itself. Now,
  suppose we want to build a new tree in the context semantics
  by extending $T$ itself. In other, word, suppose we want
  to proceed with the paths corresponding to $T$. Intuitively,
  those paths should proceed inside the box. However, we
  cannot extend $T$ itself, but subtrees of it. This
  is the reason why we extend $S_j$ and not $T$ itself
  in the last rule.
\item
  Closure conditions induced by vertices labelled with
  $C^N_\A$ can be seen as slight simplifications 
  on those induce by vertices labelled with
  $C^R_\A$. Here there are no box, we do not modify 
  the underlying stack and, accordingly, there are no 
  rules like the first two rules induced by $C^R_\A$.
\end{varitemize}
 
$T(G)$ has been defined as the smallest set satisfying certain closure
conditions. This implies it will only contain \emph{finite} trees.
Moreover, it can be endowed with an induction principle, which
does not coincide with the trivial one. For example, the first of the
two trees reported in Figure~\ref{fig:treeexamples} is smaller (as
an element of $T(G)$) than the second one, even if it is not a subtree of
it. Saying it another way, proving properties about trees $T\in T(G)$ we 
can induce on the structure of the \emph{proof} that $T$ is 
an element of $T(G)$ rather than inducing on the structure of $T$ as
a tree. This induction principle turns out to be very powerful and will
be extensively used in the following.

If $T\in T(G)$, we will denote by $U(T)$ the set containing 
all the elements of $C(G)$ which appear as third components 
of labels in $T$. The elements  $U(T)$ are the \emph{legal
stacks} for $T$. Stacks in $U(T)$ have a very constrained structure.
In particular, all vertices found (as third components of tuples)
in a legal stack are labelled with $C_\A^R$ and are precisely
the vertices of this type which lie at boundaries of boxes
in which the current edge (the second component of the tuple
labelling the root of $T$) is contained. Moreover, if a term
context $u$ and a term $s$ are found (as first and second components of tuples)
in a legal stack, then there must be a certain tree $S$ such that
$L(S)=u[s]$. More precisely: 

\begin{lemma}[Legal Stack Structure]\label{lemma:contextshape}
For every $T\in T_G$, for every $(t,e,$ $(u_1,t_1,v_1)\ldots (u_k,t_k,v_k),L)$ 
appearing as a label of a vertex of $T$:
\begin{varitemize}
  \item
  for every $i\in\{1,\ldots,k\}$ there is $f$ such that $((u_i[t_i],\rho_G(v_i),
  U_i,[\cdot]),f)\in T_G$, where $U_i=(u_{i+1},t_{i+1},v_{i+1})\ldots(u_k,t_k,v_k)$; 
  \item
  for every $i\in\{1,\ldots,k-1\}$,
  $\theta_G(v_i)=v_{i+1}$;
  \item
  $\theta_G(v_k)$ is undefined.
\end{varitemize}
Moreover, $k=0$ iff $\theta_G(e)$ is undefined and
if $k\geq 1$, then $\theta_G(e)=v_1$.
\end{lemma}
\begin{proof}
By a straightforward induction on the proof that $T\in T(G)$.\qed
\end{proof}

\subsection{Completeness}\label{subsect:completeness}
\begin{figure}[!h]
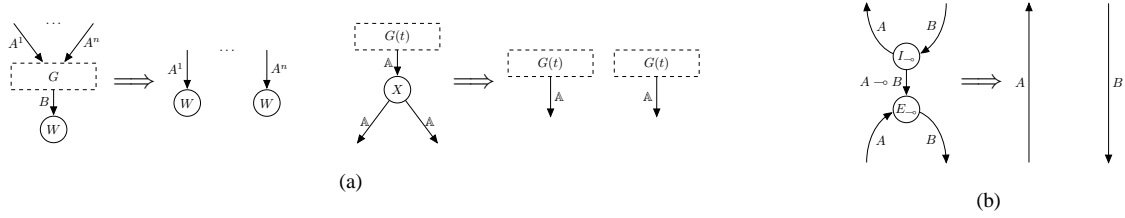

\begin{center}
\subfigure[]{
  \begin{minipage}[c]{35.5pt}
    \centering\scalebox{0.5}{\epsfbox{figure.21}}
  \end{minipage}
  \begin{minipage}[c]{20pt}
    \centering $\Longrightarrow$
  \end{minipage}
  \begin{minipage}[c]{45.5pt}
    \centering\scalebox{0.5}{\epsfbox{figure.22}}
  \end{minipage}
\hspace{20pt}
  \begin{minipage}[c]{33pt}
    \centering\scalebox{0.5}{\epsfbox{figure.36}}
  \end{minipage}
  \begin{minipage}[c]{20pt}
    \centering $\Longrightarrow$
  \end{minipage}
  \begin{minipage}[c]{73pt}
    \centering\scalebox{0.5}{\epsfbox{figure.37}}
  \end{minipage}
}
\hspace{50pt}
\subfigure[]{
  \begin{minipage}[c]{34pt}
    \centering\scalebox{0.5}{\epsfbox{figure.14}}
  \end{minipage}
  \begin{minipage}[c]{20pt}
    \centering $\Longrightarrow$
  \end{minipage}
  \begin{minipage}[c]{41.5pt}
    \centering\scalebox{0.5}{\epsfbox{figure.15}}
  \end{minipage}
}
\caption{Graph transformation produced by firing a beta-redex.}
\label{fig:betaredex}
\end{center}
\end{figure}

\begin{figure}[!h]
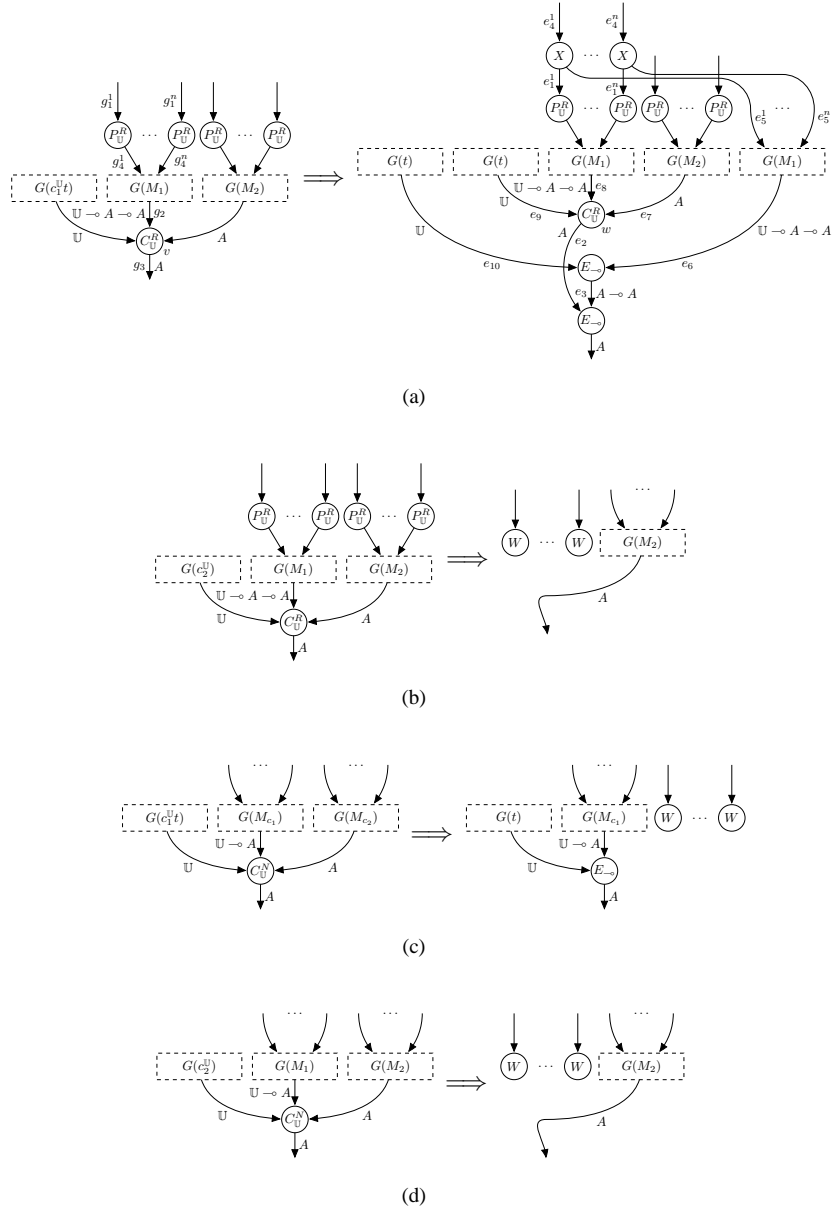

\begin{center}
  \subfigure[]{
  \begin{minipage}[c]{106pt}
    \centering\scalebox{0.5}{\epsfbox{figure.23}}
  \end{minipage}
  \begin{minipage}[c]{20pt}
    \centering $\Longrightarrow$
  \end{minipage}
  \begin{minipage}[c]{179pt}
    \centering\scalebox{0.5}{\epsfbox{figure.24}}
  \end{minipage}}\\
  \subfigure[]{
  \begin{minipage}[c]{106pt}
    \centering\scalebox{0.5}{\epsfbox{figure.40}}
  \end{minipage}
  \begin{minipage}[c]{20pt}
    \centering $\Longrightarrow$
  \end{minipage}
  \begin{minipage}[c]{70pt}
    \centering\scalebox{0.5}{\epsfbox{figure.41}}
  \end{minipage}}\\
  \subfigure[]{
  \begin{minipage}[c]{105pt}
    \centering\scalebox{0.5}{\epsfbox{figure.42}}
  \end{minipage}
  \begin{minipage}[c]{20pt}
    \centering $\Longrightarrow$
  \end{minipage}
  \begin{minipage}[c]{96pt}
    \centering\scalebox{0.5}{\epsfbox{figure.43}}
  \end{minipage}}\\
  \subfigure[]{
  \begin{minipage}[c]{105pt}
    \centering\scalebox{0.5}{\epsfbox{figure.44}}
  \end{minipage}
  \begin{minipage}[c]{20pt}
    \centering $\Longrightarrow$
  \end{minipage}
  \begin{minipage}[c]{70pt}
    \centering\scalebox{0.5}{\epsfbox{figure.45}}
  \end{minipage}}
\caption{The graph transformations induced by firing a recursive or conditional redex.}
\label{fig:recursiveconditionalredexes}
\end{center}
\end{figure}

This section is devoted to proving the completeness
of algebraic context semantics as a way to get the 
algebraic potential size of a term:
\begin{theorem}[Completeness]\label{theo:completeness}
If $\pi:\Gamma\vdash_{\h{\mathsf{A}}} M:A$, $M\leadsto^*N$
and $t$ is the argument of a redex in $N$, then
there is $T\in T(G_\pi)$
such that $L(T)=t$. 
\end{theorem}
Two lemmas will suffice for proving Theorem~\ref{theo:completeness}.
On one side, arguments of redexes inside a term $M$ can be 
retrieved in the context semantics of $M$:
\begin{lemma}[Adequacy]
If $\pi:\Gamma\vdash_{\h{\mathsf{A}}} M:A$ and $M$ contains a redex with argument $t$, 
then there is $T\in T(G_{\pi})$ such that $L(T)=t$.
\end{lemma}
\begin{proof}
First of all, we can observe that there must be a subderivation
$\xi$ of $\pi$ such that $\xi:\Delta\vdash t:\A^i$. Moreover, the
path from the root of $\pi$ to the root of $\xi$ does not
cross any instance of rule $E_\linear^R$. We can prove that 
there is $e\in E_{G_\xi}$ such that $(t,e,\varepsilon,[\cdot])\in T(G_{\xi})$
by induction on the structure of $\xi$ (with some effort if
$\A$ is not a word algebra). The thesis follows once we observe that
$G_\xi$ is a subgraph of $G_\pi$, $\theta_{G_\xi}(e)$
is always undefined and $\theta_{G_\pi}(e)$ is undefined whenever
$e$ is part of the subgraph of $G_\pi$ corresponding to $G_\xi$.
\qed
\end{proof} 
This, however, does not suffice. Context semantics must also reflect 
arguments that will eventually appear during normalization:
\begin{lemma}[Backward Preservation]
If $\pi:\Gamma\vdash_{\h{\mathsf{A}}} M:A$ and $M\leadsto N$, there is
$\xi:\Gamma\vdash N:A$ such that whenever $T\in T(G_{\xi})$,
there is $S\in T(G_{\pi})$ with $L(S)=L(T)$.
\end{lemma}
\begin{proof}
First of all we will prove the following lemma:
if $\pi:\Gamma,x:A\vdash M:B$ and $\xi:\Delta\vdash V:A$, then
the interaction graph $G_\sigma$, where
$\sigma:\Gamma,\Delta\vdash M\{V/x\}:B$ can be obtained by
plugging $G_\xi$ into the premise of $G_\pi$ corresponding
to $x$ and applying one 
or more rewriting steps as those in Figure~\ref{fig:betaredex}(a).
This lemma can be proved by an induction on
the structure of $\pi$.

Now, suppose $\pi:\Gamma\vdash M:A$ and $M\leadsto N$ by firing
a beta redex. Then, a type derivation $\xi:\Gamma\vdash N:A$ 
can be obtained from $\pi$ applying one rewriting step as that
in Figure~\ref{fig:betaredex}(b) and one or more rewriting steps
as those in Figure~\ref{fig:betaredex}(a). One can verify that,
for every rewriting step in Figure~\ref{fig:betaredex}, if $H$
is obtained from $G$ applying the rewriting step and $T\in T(H)$,
then there is $S\in T(G)$ such that $L(S)=L(T)$.

We now prove the same for conditional and recursive redexes.
To keep the proof simple, we assume to deal with conditionals
and recursion on the algebra $\U$. Suppose, 
$\pi:\Gamma\vdash M:A$ and $M\leadsto N$ by firing
a recursive redex $c_1^\U t\llan M_1,M_2\rran$. Then there is
a type derivation $\xi:\Gamma\vdash N:A$ such that $G_\xi$
can be obtained from $G_\pi$ by rewriting as in 
Figure~\ref{fig:recursiveconditionalredexes}(a). We can 
define a partial function
$$
\varphi:E_{G_\xi}\times C(G_\xi)\times\mathscr{C}_\mathscr{A}
\rightharpoonup E_{G_\pi}\times C(G_\pi)\times\mathscr{C}_\mathscr{A}
$$
in such a way that if $((t,e,U,L),f)\in T(G_\xi)$, then
there is $((t,\varphi(e,U,L)),h)\in T(G_\pi)$.
In definint $\varphi$, we will take advantage of 
Lemma~\ref{lemma:contextshape}. For example, we can assume
$U=\varepsilon$ whenever $(t,e_2,U,L)$ appear as a label
of any $T\in T(G_\xi)$. Indeed, we cannot fire any recursive
redex ``inside a box'', because the reduction relation
$\leadsto$ would forbid it.
\begin{varitemize}
\item 
  The function $\varphi$ acts as the identity on triples
  $(e,U,L)$ where $e$ lies outside the portion of $G_\xi$ affected by
  rewriting. 
\item
  Observe there are two copies of
  $G(t)$  in $G_\xi$; if $e$ is an edge of one of these two copies, then
  $\varphi(e,U,L)$ will be $(g,U,L)$, where $g$ is the edge corresponding
  to $e$ in $G(c_1^\U t)$. 
\item
  Observe there are two copies of $G(M_1)$ in $G_\xi$ , the leftmost 
  one inside a box whose premise is $w$, 
  and the rightmost outside it; if $e$ is an edge of the rightmost of these
  two copies, then $\varphi(e,U,L)$ will be $(g,U([\cdot],c_1^\U t,v),L)$,
  where $g$ is the edge corresponding to $e$ in $G(M_1)$;
  if $e$ is an edge of the leftmost of these two copies,
  then $\varphi(e,U(u,s,w),L)$ will be $(g,U(c_1^\U u,s,v),L)$.
\item
  In $G_\xi$ there is just one copy of $G(M_2)$; if $e$ is an edge of this
  copy of $G(M_2)$, then  $\varphi(e,U(u,s,w),L)$ will be 
  $(g,U(c_1^\U u,s,v),L)$.
\item
  The following equations hold:
  \begin{eqnarray*}
    \varphi(e_1^i,\varepsilon,L)&=&(g_1^i,\varepsilon,L)\\
    \varphi(e_2,\varepsilon,L)&=&(g_2,([\cdot],c_1^\U t,v),\U\linear L\linear A)\\
    \varphi(e_3,\varepsilon,A\linear L)&=&(g_3,\varepsilon,L)\\
    \varphi(e_3,\varepsilon,L\linear A)&=&(g_2,([\cdot],c_1^\U t,v),\U\linear L\linear A)
  \end{eqnarray*}
\end{varitemize}
We can prove that if $T=((r,e,\varepsilon,L),f)\in T(G_\xi)$, then
there is $((r,\varphi(e,\varepsilon,L)),h)\in T(G_\pi)$ by induction on $T$. Let us
just analyze some of the most interesting cases:
\begin{varitemize}
  \item
    Suppose there is a tree $T\in T(G_\xi)$
    whose root is labelled with $(r,e_4^i,\varepsilon,[\cdot])$. By applying
    the closure rule induced by vertices labelled with $X$, we
    can extend $T$ to a tree whose root is labelled
    with $(r,e_1^i,\varepsilon,[\cdot])$. By the induction hypothesis
    applied to $T$, there is a tree in $T(G_\pi)$
    whose root is labelled with $(r,\varphi(e_4^i,\varepsilon,[\cdot]))=(r,g_1^i,\varepsilon,[\cdot])$.
    But observe that $\varphi(e_1^i,\varepsilon,[\cdot])=(g_1^i,\varepsilon,[\cdot])$.
  \item
    Suppose there is a tree $T\in T(G_\xi)$
    whose root is labelled with $(r,e_4^i,\varepsilon,[\cdot])$. By applying
    the closure rule induced by vertices labelled with $X$, we
    can extend $T$ to a tree $S$ whose root is labelled
    with $(r,e_5^i,\varepsilon,[\cdot])$. By the induction hypothesis
    applied to $T$, there is a tree $S\in T(G_\pi)$
    whose root is labelled with $(r,\varphi(e_4^i,\varepsilon,[\cdot]))=(r,g_1^i,\varepsilon,[\cdot])$.
    By applying the closure rule induced by vertices labelled
    with $P^R_\U$, we can extend $S$ to a tree in $T(G_\pi)$
    whose root is labelled with $(r,g_4^i,([\cdot],c_1^\U t,v),[\cdot])$
    But observe that $\varphi(e_5^i,\varepsilon,[\cdot])=(g_4^i,([\cdot],c_1^\U t,v),[\cdot])$,
    because $e_5^i$ is part of the rightmost copy of $G(M_1)$.
  \item
    Suppose there is a tree $T\in T(G_\xi)$
    whose root is labelled with $(r,e_{11},\varepsilon,L)$ and $L$
    is a negative type context. By applying
    the closure rule induced by vertices labelled with $E_\linear$, we
    can extend $T$ to a tree whose root is labelled
    with $(r,e_3,\varepsilon,A\linear L)$ and, by applying again
    the same closure rule, we can obtain a tree
    whose root is labelled with $(r,e_6,\varepsilon,\U\linear A\linear L)$. By the induction hypothesis
    applied to $T$, there is a tree $S\in T(G_\pi)$
    whose root is labelled with $(r,\varphi(e_{11},\varepsilon,L))=(r,g_3,\varepsilon,L)$.
    Observe that $\varphi(e_3,\varepsilon,A\linear L)=(g_3,\varepsilon,L)$.
    By applying the closure rule induced by vertices labelled
    with $C^R_\U$, we can extend $S$ to a tree in $T(G_\pi)$
    whose root is labelled with $(r,g_2,([\cdot],c_1^\U t,v),\U\linear A\linear L)$.
    But observe that $\varphi(e_6,\varepsilon,\U\linear A\linear L)=(g_2,([\cdot],c_1^\U t,v),\U\linear A\linear L)$,
    because $e_6$ is part of the rightmost copy of $G(M_1)$.
  \item
    Suppose there is a tree $T\in T(G_\xi)$
    whose root is labelled with $(r,e_6,\varepsilon,\U\linear L\linear A)$ and $L$
    is a negative type context. By applying
    the closure rule induced by vertices labelled with $E_\linear$, we
    can extend $T$ to a tree whose root is labelled
    with $(r,e_3,\varepsilon,L\linear A)$ and, by applying another
    closure rule induced by the same vertex, we can obtain a tree
    whose root is labelled with $(r,e_2,\varepsilon,L)$. By the induction hypothesis
    applied to $T$, there is a tree $S\in T(G_\pi)$
    whose root is labelled with $(r,\varphi(e_6,\varepsilon,\U\linear L\linear A))=
    (r,g_2,([\cdot],c_1^\U t,v),\U\linear L\linear A)$.
    Observe that $\varphi(e_3,\varepsilon,L\linear A)=(g_2,([\cdot],c_1^\U t,v),\U\linear L\linear A)$
    and $\varphi(e_2,\varepsilon,L)=(g_2,([\cdot],c_1^\U t,v),\U\linear L\linear A)$.
\end{varitemize}
This shows that the thesis holds for recursive redexes
in the form $c_1^\U t\llan M_1,M_2\rran$. Similar arguments 
hold for redexes in the form $c_2^\U \llan M_1,M_2\rran$,
$c_1^\U t\llcu M_1,M_2\rrcu$,  $c_2^\U\llcu M_1,M_2\rrcu$
(see Figure~\ref{fig:recursiveconditionalredexes}(b),
Figure~\ref{fig:recursiveconditionalredexes}(c) and
Figure~\ref{fig:recursiveconditionalredexes}(d), respectively).
\qed
\end{proof}
Summing up, any possible algebraic term appearing in any possible reduct
of a typable term $M$ can be found in the context semantics
of the interaction graph for a type derivation for $M$. This proves
Theorem~\ref{theo:completeness}.
\section{On the Complexity of Normalization}\label{sect:cn}
In this section, we will give some bounds on the 
time needed to normalize terms in subsystems
$\h{\mathsf{A}}$, $\rh{\mathsf{A}}$ and $\rh{\mathsf{W}}$.
Our strategy consists in studying how constraints like linearity
and ramification induce bounds on $|L(T)|$, where $T$
is any tree built up from the context semantics. These
bounds, by Theorem~\ref{theo:completeness} and
Proposition~\ref{prop:justification}, translate into
bounds on normalization time (modulo appropriate polynomials).
Noticeably, many properties of the context semantics which
are very useful in studying $|L(T)|$ are true for all of the
above subsystems and can be proved just once. These are precisely 
the properties that that will be proved in the first part of this section.

First of all we observe
that, by definition, every subtree of $T\in T(G)$
is itself a tree in $T(G)$. Moreover, a
uniqueness property can be proved:
\begin{proposition}[Uniqueness]\label{prop:unicity}
For every interaction graph $G$, for every
$e\in E_G$, $U\in C(G)$
and $L\in\mathscr{C}_\mathscr{A}$, there is at most
one tree $T\in T(G)$ such that $T=((t,e,U,L),f)$.
\end{proposition}
\begin{proof}
We can show the following: if $((t,e,U,L),f)\in T(G)$, then
there cannot be $((s,e,U,L),g)\in T(G)$, where $s\neq t$
or $f\neq g$. We can prove this by an induction on
the structure of the proof that $((t,e,U,L),f)\in T(G)$.
First of all, observe that $L$ and $e$ uniquely determine
the last closure rule used to prove that 
$((t,e,U,L),f)\in T(G)$. In particular, if $e=(v,w)$
and $L$ is positive, then it is one induced by $v$,
otherwise it is one induced by $w$. At this point, however,
one can easily see that the domains of $f$ and $g$ must
be the same. So, there must be some $i$ such that $f(i)$
and $g(i)$ are different, but with the same label for the
root. This, however, would contraddict the inductive
hypothesis.\qed
\end{proof}
The previous result implies the following: every triple
$(e,U,L)\in E_G\times C(G)\times\mathscr{C}_\mathscr{A}$
can appear at most once in any branch of any $T\in T(G)$.
As a consequence, any $T\in T(G)$ (and, more importantly,
any $t$ such that $t=L(T)$ for some $T\in T(G)$) cannot
be too big compared to $|C(G)|$ and $|G|$. But, in turn,
the structure of relevant elements of $C(G)$ is very contrived.

Indeed, Lemma~\ref{lemma:contextshape} implies
the length of any stack in $U(T)$ where $T\in T(G_\pi)$
cannot be bigger than the recursion depth $R(\pi)$ of
$\pi$: the length of $U$ is equal to the ``depth''
of $e$ whenever $(t,e,U,L)$ appears as a label in $T$.

Along a path, the fourth component of the underlying
tuple can change, but there is something which stays
invariant:
\begin{lemma}\label{lemma:guiding}
For every $T\in T(G)$ there is a type $\A^i$ such that
for every $(t,e,U,L)$ appearing as a label of a vertex
of $T$, $\beta_G(e)=L[\A^i]$. We will say $T$ is
\emph{guided} by $\A^i$.
\end{lemma} 
\begin{proof}
By a straigthforward induction on the proof that $T\in T(G)$.\qed
\end{proof}
The previous lemmas shed some light on the
combinatorial properties of tuples 
$(t,e,U,L)\in S(G_\pi)$ labelling vertices of trees in $T(G_\pi)$.
This is enough to prove $|L(T)|$ to be exponentially related to the 
cardinality of $U(T)$: 
\begin{proposition}\label{prop:boundfirst}
Suppose $\pi:\Gamma\vdash_{\h{\mathsf{A}}} M:A$
and $T\in T(G_\pi)$. Then $|L(T)|\leq \mathscr{K}_{\mathscr{A}}^{|G_\pi||U(T)|}$.
\end{proposition}
\begin{proof}
First of all, we observe that whenever $(t,e,U,L)$ labels
a vertex $v$ of $T$ and $(s,f,V,M)$ labels one child of $v$,
then either $s=t$ or $t=cs_1\ldots s_k$ and $L=\A\linearo{k}[\cdot]$.
The thesis follows from lemmas~\ref{lemma:contextshape} and~\ref{lemma:guiding}.\qed 
\end{proof}
This will lead to prove primitive recursive bounds for $\h{\mathsf{A}}$ and elementary
bounds for $\rh{\mathsf{A}}$. However, we cannot expect to prove any polynomial bound
from Proposition~\ref{prop:boundfirst}. In the case of $\rh{\mathsf{W}}$, 
a stronger version of Proposition~\ref{prop:boundfirst} can be proved by exploiting
ramification.
\begin{proposition}\label{prop:boundfirstramified}
Suppose $\pi:\Gamma\vdash_{\rh{\mathsf{W}}} M:A$
and $T\in T(G_\pi)$. Then $|L(T)|\leq |G_\pi||U(T)|$.
\end{proposition}
\begin{proof}
First of all, we prove the following lemma:
for every $T\in T(G_\pi)$, if $T$ is guided by
$\A^i$ and $\A$ is a word algebra, there are at most
one tree $((t,e,U,L),f)\in T(G_\pi)$ and one integer
$i\in\N$ such that
$f(i)=T$. To prove the lemma, suppose
$((t,e,U,L),f),((s,g,V,M),h)\in T(G_\pi)$ and
$f(i)=g(j)=T$. $T$ uniquely determines the
closure condition used to prove that both
$((t,e,U,L),f)$, $((s,g,V,M),h)\in T(G_\pi)$,
which must be the same because those induced
by typing rule $X$ are forbidden. But by inspecting
all the closure rules, we can conclude that 
$s=t$, $e=g$, $U=V$, $L=M$ and $f(n)=g(n)$ 
for every $n$. Then, we can proceed exactly
as in Proposition~\ref{prop:boundfirst}.\qed
\end{proof}
Notice how the elementary bound of Proposition~\ref{prop:boundfirst}
has become a polynomial bound in Proposition~\ref{prop:boundfirstramified}.
Quite surprisingly, this phase transition happens as soon as the
class of types on which we allow contraction is restricted from
$\mathsf{A}$ to $\mathsf{W}$.
\subsection{$\h{\mathsf{A}}$ and Primitive Recursion}
Given an interaction graph $G$, we now need to define subclasses $T(\mathcal{U})$ of
$T(G)$ for any subset $\mathcal{U}$ of $C(G)$. In principle, we would like
$U(T)$ to be a subset of $\mathcal{U}$ whenever $T\in T(\mathcal{U})$. However, this
is too strong a constraint, since we should allow $U(T)$ to contain extensions
of stacks in $\mathcal{U}$, the extensions being obtained themselves in this constrained way.
The following definition captures the above intuition.
Let $G$ be an interaction graph and $\mathcal{U}\subseteq C(G)$.
A tree $T\in T(G)$ is said
to be \emph{generated} by $\mathcal{U}$ iff for
every $U\in U(T)$:
\begin{varitemize}
  \item
  either $U\in\mathcal{U}$,
  \item
  or $U=(u_1,t_1,v_1)\ldots(u_k,t_k,v_k)V$ where $V$ is itself in $\mathcal{U}$
  and has maximal length (between all the elements of $\mathcal{U}$. Moreover, 
  for every $i\in\{1,\ldots,k\}$, the  tree $((u_i[t_i],\rho_G(v_i),
  (u_{i+1},\allowbreak t_{i+1},v_{i+1})\ldots(u_k,t_k,v_k)V,\allowbreak[\cdot]),f)$
  must be itself generated by $\mathcal{U}$.
\end{varitemize}
The set of all trees generated by $\mathcal{U}$ will be denoted
by $T(\mathcal{U})$. This definition is well-posed because of the induction principles on $T(G)$. 
Indeed, we require some trees $T_1,\ldots,T_n$ to be in $T(\mathcal{U})$
when defining conditions on $T$ being an element of $T(\mathcal{U})$ itself;
however, $T_1,\ldots,T_n$ are ``smaller'' than $T$. Notice that $T$ is not monotone
as an operator on subsets of $C(G)$. For example, $T(\{\varepsilon\})=T(G)$,
while $T(\{\varepsilon,C\})\subset T(G)$ whenever $C\notin U(T)$ for any
$T\in T(G)$. This is due to the requirement of $V$ having maximal length
in the definition above.

\begin{lemma}
For every $d\in\N$ there is a primitive recursive function
$p_d:\N^2\rightarrow\N$ such that if $\pi:\Gamma\vdash_{\h{\mathsf{A}}} M:A$,
$\mathcal{U}\subseteq C(G_\pi)$, the maximal length of elements of $\mathcal{U}$
is $n$, and $T\in T(\mathcal{U})$, then 
$|L(T)|\leq p_{R(\pi)-n}(|G_\pi|,|\mathcal{U}|)$.
\end{lemma}
\begin{proof}
We can put
\begin{eqnarray*}
p_0(x,y)&=&\mathscr{K}_\mathscr{A}^{xy}\\
\forall i\geq 1.h_i(x,y,0)&=&\mathscr{K}_\mathscr{A}^{xy}\\
\forall i\geq 1.h_i(x,y,z+1)&=&h_i(x,y,z)+p_{i-1}(x,y+h_i(x,y,z))\\
\forall i\geq 1.p_i(x,y)&=&h_i(x,y,xy)
\end{eqnarray*}
Every $p_i$ and $h_i$ are primitive recursive. Moreover, all
these functions are monotone in each of their arguments.
We will now prove the
thesis by induction on $R(\pi)-n$. If $R(\pi)=n$, then there are
elements in $\mathcal{U}$ having length equals to $R(\pi)$. This,
by Lemma~\ref{lemma:contextshape} and Proposition~\ref{prop:boundfirst}, 
implies that if $T\in T(G)$ is
generated by $\mathcal{U}$, then
$|L(T)|$ is bounded by $p_0(|G|,|\mathcal{U}|)$ since
none of the elements of $\mathcal{U}$ having maximal
length can be extended into an element of $U(T)$
and, as a consequence, $U(T)\subseteq\mathcal{U}$.
Now, let us suppose $R(\pi)-n\geq 1$. Let us define
$\mathcal{W}\subseteq C(G)$ as follows 
$$
\mathcal{W}=\{(u,t,v)U\spb U\in\mathcal{U}\mbox{ has maximal length and }((u[t],\rho_G(v),U,[\cdot]),f)\in T(\mathcal{U})\}
$$
Clearly, $T(\mathcal{U}\cup\mathcal{W})=T(\mathcal{U})$.
Now, consider the sequence $(v_1,U_1),\ldots,(v_k,U_k)$ of all the
pairs $(v_i,U_i)\in V_G\times\mathcal{U}$ such that $(u,t,v_i)U_i\in\mathcal{W}$
for some $u,t$. Obviously, $k\leq |G||\mathcal{U}|$. 
If $k=0$, then the thesis is trivial, since
\begin{eqnarray*}
|L(T)|&\leq& \mathscr{K}_\mathscr{A}^{|G||\mathcal{U}|}\\
  &=& h_{R(\pi)-n}(|G|,|\mathcal{U}|,0)\\
  &\leq& h_{R(\pi)-n}(|G|,|\mathcal{U}|,|G||\mathcal{U}|)\\
  &=& p_{R(\pi)-n}(|G|,|\mathcal{U}|).
\end{eqnarray*}
From now on, suppose $k\geq 1$. Let $\mathcal{W}_1,\ldots,\mathcal{W}_k\subseteq\mathcal{W}$
be defined as follows: $\mathcal{W}_i=\{(u,t,v_j)U_j\in\mathcal{W}\spb j\leq i\}$.
By definition, $\mathcal{W}_k=\mathcal{W}$.
We can assume, without losing generality, that 
\begin{varitemize}
  \item
  $T_1=((t_1,\rho_G(v_1),U_1,[\cdot]),f_1)$ only contains elements from $\mathcal{U}$
  as part of its labels.
  \item
  For every $i\in\{2,\ldots,k\}$,
  the tree $T_i=((t_i,\rho_G(v_i),U_i,[\cdot]),f_i)$ is generated
  by $\mathcal{U}\cup\mathcal{W}_{i-1}$. 
\end{varitemize}
We can now prove that
$$
\sum_{j=1}^{i+1}|t_j|\leq h_{R(\pi)-n}(|G|,|\mathcal{U}|,i)
$$
by induction on $i$. The tree $T_1$
only contains elements of $\mathcal{U}$ as part of
its labels and, by Proposition~\ref{prop:boundfirst},
$$
|t_1|\leq\mathscr{K}_\mathscr{A}^{|G||U(T_1)|}\leq\mathscr{K}_\mathscr{A}^{|G||\mathcal{U}|}= h_{R(\pi)-n}(|G|,|\mathcal{U}|,0).
$$
If $i\geq 1$, by inductive hypothesis (on $i$) we get
$$
\sum_{j=1}^i|t_j|\leq h_{R(\pi)-n}(|G|,|\mathcal{U}|,i-1).
$$
This hields $|\mathcal{W}_i|\leq h_{R(\pi)-n}(|G|,|\mathcal{U}|,i-1)$,
because for every every term $t_j$ there at most $|t_j|$ triples
$(u,s,v_j)$ such that $u[s]=t_j$. By induction hypothesis
(both on $i$ and $R(\pi)-n$), we get
\begin{eqnarray*}
\sum_{j=1}^{i+1}|t_j|&=&\sum_{j=1}^i|t_j|+|t_{i+1}|\\
  &\leq&h_{R(\pi)-n}(|G|,|\mathcal{U}|,i-1)+
  p_{R(\pi)-n-1}(|G|,|\mathcal{U}|+h_{R(\pi)-n}(|G|,|\mathcal{U}|,i-1))\\
  &=&h_{R(\pi)-n}(|G|,|\mathcal{U}|,i),
\end{eqnarray*} 
because $T_{i+1}$ is generatged by $\mathcal{U}\cup\mathcal{W}_{i-1}$.
So, $|\mathcal{W}|=|\mathcal{W}_k|\leq h_{R(\pi)-n}(|G|,|\mathcal{U}|,k-1)$.
Now, suppose $T\in T(\mathcal{U})=T(\mathcal{U}\cup\mathcal{W})$. 
Then by inductive hypothesis (on $R(\pi)-n$)
\begin{eqnarray*}
|L(T)|&\leq& p_{R(\pi)-n-1}(|G|,|\mathcal{U}\cup\mathcal{W}|)\\
  &=& p_{R(\pi)-n-1}(|G|,|\mathcal{U}|+|\mathcal{W}|)\\
  &=& p_{R(\pi)-n-1}(|G|,|\mathcal{U}|+|\mathcal{W}_k|)\\
  &\leq& p_{R(\pi)-n-1}(|G|,|\mathcal{U}|+h_{R(\pi)-n}(|G|,|\mathcal{U}|,k-1))\\
  &\leq& h_{R(\pi)-n}(|G|,|\mathcal{U}|,k)\\
  &\leq& h_{R(\pi)-n}(|G|,|\mathcal{U}|,|G||\mathcal{U}|)\\
  &=& p_{R(\pi)-n}(|G|,|\mathcal{U}|).
\end{eqnarray*}
This concludes the proof.\qed
\end{proof}
As a corollary, we get:
\begin{theorem}\label{theo:soundra}
For every $d\in\N$, there is a primitive recursive 
function $p_d:\N\rightarrow\N$ such that for every type derivation 
$\pi:\Gamma\vdash_{\mathbf{H(A)}} M:A$, if 
$T\in T(G_\pi)$
then $|L(T)|\leq p_{R(\pi)}(|M|)$.
\end{theorem}
\begin{proof}
Trivial, since every tree $T\in T(G_\pi)$ is
generated by $\{\varepsilon\}$.\qed
\end{proof}
Theorem~\ref{theo:soundra} implies, by Proposition~\ref{prop:justification},
that the time needed to normalize a term $M$ with a type derivation $\pi$
in $\h{\mathsf{A}}$ is bounded by a primitive recursive function
(just depending on the recursion depth of $\pi$) applied to 
the size of $M$. This, in particular, implies that every 
function $f:\N\rightarrow\N$ which can be represented in $\h{\mathsf{A}}$
must be primitive recursive, because all terms corresponding to
calls to $f$ can be typed with bounded-recursion-depth type derivations.
This is a \emph{leitmotif}: elementary bounds for $\rh{\mathsf{A}}$
and polynomial bounds for $\rh{\mathsf{W}}$ will have the same flavour. 
Observe how this way of formulating soundness results is
necessary in a higher-order setting. Indeed, since bounds are given on the normalization
time of \emph{any} term in the subsystem and the subsystem itself is complete
for a complexity class, we cannot hope to prove, say, that any
term in $\h{\mathsf{A}}$ can be normalized with a fixed, primitive recursive,
bound on its size.
\subsection{$\rh{\mathsf{A}}$ and Elementary Time}\label{sect:rha}
Now, consider the interpretation of branches of trees in $T(G)$ 
as paths in $G$: any such path can enter and exit boxes by traversing
vertices labelled with $C^R_\A$ or $P^R_\A$. The stack $U$ in the underlying
context can change as a result of the traversal. Indeed, $U$ changes only
when entering and exiting boxes (other vertices of $G$ leave $U$ unchanged,
as can be easily verified). As a consequence, by Proposition~\ref{prop:boundfirst}, 
entering and exiting boxes is essential to obtain a hyperexponential complexity: if
paths induced by a tree $T\in T(G)$ do not enter or exit boxes, $U(T)$ will
be a singleton and $L(T)$ will be bounded by a fixed exponential on $|G|$.
In general, paths induced by trees can indeed enter or exit boxes.
If ramification holds, on the other hand, a path induced by a tree
guided by $\A^i$ entering into a box whose main premise is labelled
by $\F^j$ (where $j\leq i$), will stay inside the box and the third component
of the underlying context will only increase in size. More formally:
\begin{lemma}\label{lemma:ramified}
Suppose $\pi$ to be a type derivation satisfying the ramification
condition, $S\in T(G_\pi)$ to be guided by $\A^i$,
$(t,e,U,L)$ to label a vertex $v$ of $S$ and $U=(u,t,w)V$, 
where $\beta_G(\rho_G(w))=\F^j$ and $j\leq i$. Then all the ancestors
of $v$ in $S$ are labelled with quadruples $(s,f,W,M)$ where $W=ZU$.  
\end{lemma}
\begin{proof}
By a straightforward induction on the structure of $S$. 
In particular, the only vertices in $G_\pi$ whose closure conditions
affect the third component of $C(G)$ are those labelled with
$P^R_\G$ and $C^R_\G$, where $\G$ is any free algebra. The rule induced
by a vertex $P^R_\G$, however, makes the underlying stack
bigger (from $U$, it becomes $(u,t,v)U$). As a consequence, the statment
of the lemma is verified. Now, consider rules induced by $C^R_\G$
vertices:
\begin{varitemize}
  \item
    The first four rules cannot be applied under this lemma's hypothesis:
    by ramification $V(B_0)>V(A)$ but this is in contraddiction with
    $j\leq i$ from lemma's hypothesis.
  \item
    The fifth rule is a bit delicate: $S_j$ satisfies the lemma, being 
    it a subtree of a tree $T$ to which we can apply the inductive hypothesis.
    The rule appens a node whose third component is $(u,s,v)U$, where
    $U$ is the third component of the tuple labelling the root of $T$.
    The thesis clearly holds.
\end{varitemize}
This concludes the proof. \qed
\end{proof}
This in turn allows to prove a theorem bounding the 
algebraic potential size of terms in system $\rh{\mathsf{A}}$:
\begin{theorem}\label{theo:soundnessrha}
For every $d,e\in\N$, there are elementary functions 
$p^d_e:\N\rightarrow\N$ such
that for every type derivation 
$\pi:\Gamma\vdash_{\rh{\mathsf{A}}} M:A$, if 
$T\in T(G_\pi)$
then $|L(T)|\leq p^{I(\pi)}_{R(\pi)}(|M|)$.
\end{theorem}
\begin{proof}
Consider the following elementary functions:
\begin{eqnarray*}
\forall n,m\in\N.p^n_m&:&\N\rightarrow\N;\\
p^0_{m}(x)&=&\mathscr{K}_\mathscr{A}^{x^2};\\
p^{n+1}_{m}(x)&=&\mathscr{K}_\mathscr{A}^{x(x\cdot p^n_m(x))^m}.
\end{eqnarray*}
First of all, notice that for every $x,m,n$, $p^{n+1}_{m}(x)\geq p^n_m(x)$.
We will prove that if $T=((t,e,U,L),f)\in T(G_\pi)$
is guided by $\A^i$, then
$|t|\leq p^j_{R(\pi)}(|G_\pi|)$,
where $j=\max\{I(\pi)-i,0\}$. We go by induction
on $j$. 

If $j=0$, then $I(\pi)\leq i$. This implies
that $|U(T)|\leq|G_\pi|$, by 
lemmas~\ref{lemma:contextshape} and~\ref{lemma:ramified}.
Indeed, by Lemma~\ref{lemma:ramified} stacks can only get bigger
along paths induced by $T$ and any vertex in $G$ uniquely
determines the length of stacks (Lemma~\ref{lemma:contextshape}).
As a consequence, $|t|\leq \mathscr{K}^{|G_\pi|^2}_{\mathscr{A}}=p_{R(\pi)}^0(|G_\pi|)$. 

Now, suppose the thesis holds for $j$ and suppose
$T=((t,e,U,L),f)$ to be guided by $\A^i$, where $I(\pi)-i=j+1$. 
By Lemma~\ref{lemma:ramified} and the induction hypothesis,
$|U(T)|\leq(|G_\pi|p_{R(\pi)}^j(|G_\pi|))^{R(\pi)}$. Indeed, elements 
of $U(T)$ are stacks in the form $(u_1,t_1,v_1)\cdots(u_k,t_k,v_k)$
where $k\leq R(\pi)$ and, for every $l\in\{1,\ldots,k\}$:
\begin{varitemize}
  \item
    Either $\beta_G(\rho_G(v_l))=\F^h$, where $h\leq i$ and $T$ uniquely determines
    $u_l$ and $t_l$ due to Lemma~\ref{lemma:ramified};
  \item
    or $\beta_G(\rho_g(v_l))=\F^h$ where $h>l$,
    $(u_{l+1},t_{l+1},v_{l+1})\cdots(u_k,t_k,v_k)$ and $v_l$ uniquely determines
    $u_l[t_l]$ and $|u_l[t_l]|\leq p_{R(\pi)}^j(|G_\pi|)$ by the
    inductive hypothesis.
\end{varitemize}
As a consequence,
$$
|t|\leq\mathscr{K}^{|G_\pi|(|G_\pi|p_{R(\pi)}^j(|G_\pi|))^{R(\pi)}}_\mathscr{A}
\leq p_{R(\pi)}^{j+1}(|G_\pi|).
$$
The thesis follows by observing that $j\leq I(\pi)$.
\qed
\end{proof}
This implies that every function which can be represented inside
$\rh{A}$ is elementary time computable.
\subsection{$\rh{\mathsf{W}}$ and Polynomial Time}
Notice that the exponential bound of
Proposition~\ref{prop:boundfirst} has
become a polynomial bound in Proposition~\ref{prop:boundfirstramified}.
Since Proposition~\ref{prop:boundfirst} has been the essential ingredient in proving
the elementary bounds of Section~\ref{sect:rha}, polynomial bounds are
to be expected for $\rh{\mathsf{W}}$. Indeed:
\begin{theorem}\label{theo:soundnessrhw}
For every $d,e\in\N$, there are polynomials 
$p^d_e:\N\rightarrow\N$ such
that for every type derivation 
$\pi:\Gamma\vdash_{\rh{\mathsf{W}}} M:A$, if 
$T\in T(G_\pi)$ then $|L(T)|\leq p^{I(\pi)}_{R(\pi)}(|M|)$.
\end{theorem}
\begin{proof}
We can proceed very similarly to the proof of Theorem~\ref{theo:soundnessrha}.
Consider the following polynomials:
\begin{eqnarray*}
\forall n,m\in\N.p^n_m&:&\N\rightarrow\N\\
p^0_{m}(x)&=&x^2\\
p^{n+1}_{m}(x)&=&x(x\cdot p^n_m(x))^n
\end{eqnarray*}
For every $x,m,n$, $p^{n+1}_{m}(x)\geq p^n_m(x)$.
We will prove that if $T=((t,e,U,L),f)\in T(G_\pi)$
is generated by $\A^i$, then
$|t|\leq p^j_{R(\pi)}(|G_\pi|)$,
where $j=\max\{I(\pi)-i,0\}$. We go by induction
on $j$. 

If $j=0$, then $I(\pi)\leq i$. This implies
that $|U(T)|\leq|G_\pi|$ by 
lemmas~\ref{lemma:contextshape} and \ref{lemma:ramified},
similarly as in Theorem~\ref{theo:soundnessrha}.
As a consequence, $|t|\leq |G_\pi|^2=p_{R(\pi)}^0(|G_\pi|)$. 

Now, suppose the thesis holds for $j$ and suppose
$T=((t,e,U,L),f)$ to be guided by $\A^i$, where $R(\pi)-i=j+1$. 
By Lemma~\ref{lemma:ramified} and the induction hypothesis,
$|U(T)|\leq(|G_\pi|p_{R(\pi)}^j(|G_\pi|))^{R(\pi)}$, similarly
as in theorem~\ref{theo:soundnessrha}. As a consequence,
$$
|t|\leq|G_\pi|(|G_\pi|p_{R(\pi)}^j(|G_\pi|))^{R(\pi)}\leq p_{R(\pi)}^{j+1}(|G_\pi|))
$$
The thesis follows by observing that $j\leq I(\pi)$.
\qed
\end{proof}

\section{Embedding Complexity Classes}\label{sect:completeness}
In this section, we will provide embeddings of
$\mathbf{FR}$ into $\mathsf{H(\emptyset)}$,
$\mathbf{FE}$ into $\mathsf{RH(A)}$
and $\mathbf{FP}$ into $\mathsf{RH(\emptyset)}$.
This will complete the picture sketched 
in Section~\ref{sect:subsystems}. First of all,
we can prove that a weaker notion of contraction
can be retrieved even if $\mathbf{D}=\emptyset$:
\begin{lemma}\label{lemma:contr}
For every term $M$, there is a term $[M]_{x,y}^w$ such
that for every $t\in\mathscr{E}_\U$,
$([M]_{x,y}^w)\{t/w\}\leadsto^*M\{t/x,t/y\}$.
For every $n\in\N$, if $\Gamma,x:\U^n,y:\U^n\vdash_{\mathbf{H(\emptyset)}}M:A$
then $\Gamma,w:\U^n\vdash_{\mathbf{H(\emptyset)}}[M]_{x,y}^w:A$ and
if $\Gamma,x:\U^n,y:\U^n\vdash_{\rh{\emptyset}}M:A$
then $\Gamma,w:\U^{n+1}\vdash_{\rh{\emptyset}}[M]_{x,y}^w:A$.
\end{lemma}
\begin{proof}
Given a term $t\in\mathscr{E}_\U$, the term
$\overline{t}\in\mathscr{E}_\C$ is defined
as follows, by induction on $t$:
\begin{eqnarray*}
\overline{c_2^\U}&=&c_2^\C;\\
\overline{c_1^\U t}&=& c_1^\C\overline{t}c_2^\C.
\end{eqnarray*}
We can define two closed terms $\Duplicate,
\Extract\in\mathscr{M}_\mathscr{A}$ such that, 
for every $t\in\mathscr{E}_\U$
\begin{eqnarray*}
\Extract\;\overline{t}&\leadsto^*&t;\\
\Duplicate\;t&\leadsto^*& c_2^\C \overline{t}\;\overline{t}. 
\end{eqnarray*}
The terms we are looking for are the following:
\begin{eqnarray*}
\Extract&\equiv&\lambda x.x\llan\lambda y.\lambda w.\lambda z.\lambda q.c_1^\U z,c_2^\U\rran;\\
\Duplicate&\equiv&\lambda x.x\llan
                    \lambda y.\lambda w.w\llcu\lambda z.\lambda q.c_1^\C(c_1^\C z c_2^\C)(c_1^\C q c_2^\C),c_2^\C\rrcu,
                    c_1^\C c_2^\C c_2^\C\rran.
\end{eqnarray*}
Indeed:
\begin{eqnarray*}
c_2^\C\llan\lambda y.\lambda w.\lambda z.\lambda q.c_1^\U z,c_2^\U\rran&\leadsto&c_2^\U;\\
  c_1^\C\overline{t}c_2^\C\llan\lambda y.\lambda w.\lambda z.\lambda q.c_1^\U z,c_2^\U\rran
  &\leadsto^*&(\lambda y.\lambda w.\lambda z.\lambda q.c_1^\U z)\;\overline{t}\;c_2^\C\\
  &&(\overline{t}\llan\lambda y.\lambda w.\lambda z.\lambda q.c_1^\U z,c_2^\U\rran)\;c_2^\U\\
  &\leadsto^*&(\lambda y.\lambda w.\lambda z.\lambda q.c_1^\U z)\;\overline{t}\;c_2^\C\;t\;c_2^\U\\
  &\leadsto^*& c_1^\U t;\\
\Extract\;\overline{t}&\leadsto&\overline{t}\llan\lambda y.\lambda w.\lambda z.\lambda q.c_1^\U z,c_2^\U\rran\\
  &\leadsto^*& t;\\
c_2^\U\llan\lambda y.\lambda w.w\llcu\lambda z.\lambda q.c_1^\C(c_1^\C z c_2^\C)(c_1^\C q c_2^\C),c_2^\C\rrcu,
  c_1^\C c_2^\C c_2^\C\rran&\leadsto^*& c_1^\C c_2^\C c_2^\C;\\
c_1^\U\;t\llan\lambda y.\lambda w.w\llcu\lambda z.\lambda q.c_1^\C(c_1^\C z c_2^\C)(c_1^\C q c_2^\C),c_2^\C\rrcu,
  c_1^\C c_2^\C c_2^\C\rran&\leadsto^*&c_1^\C\overline{t}\;\overline{t}
  \llcu\lambda z.\lambda q.c_1^\C(c_1^\C z c_2^\C)(c_1^\C q c_2^\C),c_2^\C\rrcu\\
  &\leadsto^*& c_1^\C(c_1^\C \overline{t} c_2^\C)(c_1^\C \overline{t} c_2^\C);\\
\Duplicate\;t&\leadsto&t\llan\lambda y.\lambda w.w\llcu\lambda z.\lambda q.c_1^\C(c_1^\C z c_2^\C)(c_1^\C q c_2^\C),\\
  &&c_2^\C\rrcu,c_1^\C c_2^\C c_2^\C\rran\\
  &\leadsto^*&c_1^\C\overline{t}\;\overline{t}.
\end{eqnarray*}
Observe that, for every natural number $n$:
\begin{eqnarray*}
\vdash_{\mathbf{H(\emptyset)}}:\Extract&:&\C^n\linear\U^n;\\
\vdash_{\mathbf{H(\emptyset)}}:\Duplicate&:&\U^n\linear\C^n;\\
\vdash_{\mathbf{H(\emptyset)}}:\Extract&:&\C^{n+1}\linear\U^n;\\
\vdash_{\mathbf{H(\emptyset)}}:\Duplicate&:&\U^{n+1}\linear\C^n.
\end{eqnarray*}
Now let us define:
$$
[M]_{x,y}^w\equiv(\Duplicate\;w)\llcu\lambda z.\lambda q.
  (\lambda x.\lambda y.M)(\Extract\;z)(\Extract\;q),\lambda x.\lambda y.M\rrcu
$$
Indeed, for every $t\in\mathscr{E}_\U$:
\begin{eqnarray*}
[M]_{x,y}^w\{t/w\}&\equiv&(\Duplicate\;t)\llcu\lambda z.\lambda q.
  (\lambda x.\lambda y.M)(\Extract\;z)(\Extract\;q),\lambda x.\lambda y.M\rrcu\\
&\leadsto^*& (c_1^\C \overline{t}\overline{t})\llcu\lambda z.\lambda q.
  (\lambda x.\lambda y.M)(\Extract\;z)(\Extract\;q),\lambda x.\lambda y.M\rrcu\\
&\leadsto^*& (\lambda x.\lambda y.M)(\Extract\;\overline{t})(\Extract\;\overline{t})\\
&\leadsto^*& (\lambda x.\lambda y.M)\;t\;t\\
&\leadsto^*& M\{t/x,t/y\}.
\end{eqnarray*}
Observe that the requirement of typings for $[M]_{x,y}^w$ can be easily verified.\qed
\end{proof}
The above lemma suffices to prove every primitive recursive
function to be representable inside $\h{\emptyset}$:
\begin{theorem}
For every primitive recursive function $f:\N^n\rightarrow\N$ there
is a term $M_f$ such that $\vdash_{\h{\emptyset}} M_f:\U^0\linearo{n}\U^0$
and $M_f$ represents $f$.
\end{theorem}
\begin{proof}
Base functions are the constant $0:\N\rightarrow\N$, the successor
$s:\N\rightarrow\N$ and for every $n,i$ projections
$u^n_i:\N^n\rightarrow\N$. It can be easily checked that these
functions are represented by
\begin{eqnarray*}
M_0&\equiv&\lambda x.c_2^\U;\\
M_s&\equiv&\lambda x.c_1^\U x;\\
M_{u^n_i}&\equiv&\lambda x_1.\lambda x_2.\ldots.\lambda x_n.x_i.
\end{eqnarray*}
Observe that
\begin{eqnarray*}
\vdash_{\mathbf{H(\emptyset)}}M_0&:&\U^0\linear\U^0;\\
\vdash_{\mathbf{H(\emptyset)}}M_s&:&\U^0\linear\U^0;\\
\vdash_{\mathbf{H(\emptyset)}}M_{u^n_i}&:&\U^0\linearo{n}\U^0.
\end{eqnarray*}
We now need some additional notation. Given a term $M$
and $n$ variables $x_1,\ldots,x_n$, we will define
terms $M_i^{x_1,\ldots,x_n}$ as follows:
\begin{eqnarray*}
M_1^{x_1,\ldots,x_n}&\equiv&(\lambda x_1.\ldots.\lambda x_n.M)x_1;\\
\forall i\geq 1.M_{i+1}^{x_1,\ldots,x_n}&\equiv&[M_i^{x_1,\ldots,x_n} x_{i+1}]_{x_i,x_{i+1}}^{x_{i+1}}.
\end{eqnarray*}
We can prove the following by induction on $i$:
$$
M_i^{x_1,\ldots,x_n}\{t/x_i\}\leadsto^*(\lambda x_1.\ldots.\lambda x_n.M)\underbrace{t\ldots t}_{\mbox{$i$ times}}.
$$
Indeed:
\begin{eqnarray*}
M_1^{x_1,\ldots,x_n}\{t/x_i\}&\equiv&(\lambda x_1.\ldots.\lambda x_n.M)t;\\
\forall i\geq 2.M_{i+1}^{x_1,\ldots,x_n}\{t/x_{i+1}\}&\equiv&[M_i^{x_1,\ldots,x_n}x_{i+1}]_{x_i,x_{i+1}}^{x_{i+1}}\{t/x_{i+1}\}\\
  &\leadsto^*&(M_i^{x_1,\ldots,x_n}x_{i+1})\{t/x_i,t/x_{i+1}\}\\
  &\leadsto^*&(M_i^{x_1,\ldots,x_n}\{t/x_i\})t\\
  &\leadsto^*&((\lambda x_1.\ldots.\lambda x_n.M)\underbrace{t\ldots t}_{\mbox{$i$ times}})t\\
  &\equiv&(\lambda x_1.\ldots.\lambda x_n.M)\underbrace{t\ldots t}_{\mbox{$i+1$ times}}.
\end{eqnarray*}
In this way we can get a generalized variant of Lemma~\ref{lemma:contr} by 
putting $\langle M\rangle_{x_1,\ldots,x_n}^z\equiv (\lambda x_n.M_n^{x_1,\ldots,x_n})z$.
Indeed:
\begin{eqnarray*}
\langle M\rangle_{x_1,\ldots,x_n}^z\{t/z\}&\equiv& (\lambda x_n.M_n^{x_1,\ldots,x_n})t\\
  &\leadsto&M_n^{x_1,\ldots,x_n}\{t/x_n\}\\
  &\leadsto^*&(\lambda x_1.\ldots.\lambda x_n.M)\underbrace{t\ldots t}_{\mbox{$n$ times}}\\
  &\leadsto^*& M\{t/x_1,\ldots t/x_n\}.
\end{eqnarray*}
We are now ready to prove that composition and recursion can be represented
in $\mathsf{H}(\emptyset)$. Suppose $f:\N^n\rightarrow\N$, $g_1,\ldots,g_n:\N^m\rightarrow\N$
and let $h:\N^m\rightarrow\N$ be the function obtained by composing $f$ with $g_1,\ldots,g_n$,
i.e.
$$
h(n_1,\ldots,n_m)=f(g_1(n_1,\ldots,n_m),\ldots,g_n(n_1,\ldots,n_m)).
$$
We define
\begin{eqnarray*}
N&\equiv& \lambda x_1^m.\ldots.\lambda x_n^m.\dots.\lambda x_1^1.\ldots.\lambda x_n^1.
  M_f(M_{g_1}x_1^1\ldots x_1^m)\ldots(M_{g_n}x_n^1\ldots x_n^m);\\
M_h^m&\equiv&\langle N x_1^m \ldots x_n^m \rangle^{y_m}_{x_1^m,\ldots,x_n^m};\\
\forall i<m.M_h^i&\equiv& \langle \lambda y_{i+1}.(M_h^{i+1}x_1^{i} \ldots x_n^{i})\rangle^{y_i}_{x_1^i,\ldots,x_n^i};\\
M_h&\equiv&\lambda y_1.M_h^1.
\end{eqnarray*}
Indeed:
\begin{eqnarray*}
M_h \inttou{n_1}\ldots \inttou{n_m}&\leadsto^*& (M_h^1\{\inttou{n_1}/y_1\})\inttou{n_2}\ldots \inttou{n_m}\\
  &\leadsto^*& (\lambda y_{2+1}.M_h^2 \inttou{n_1}\ldots \inttou{n_1})\inttou{n_2}\ldots \inttou{n_m}\\
  &\leadsto& (M_h^2\{\inttou{n_2}/y_2\} \inttou{n_1}\ldots \inttou{n_1})\inttou{n_3}\ldots \inttou{n_m}\\
  &\leadsto^*&\ldots\\
  &\leadsto^*&(\ldots((N \inttou{n_m}\ldots \inttou{n_m})\inttou{n_{m-1}}\ldots\inttou{n_{m-1}})\ldots)\inttou{n_1}\ldots \inttou{n_1}\\
  &\leadsto^*& M_f(M_{g_1}\inttou{n_1}\ldots \inttou{n_m})\ldots(M_{g_n}\inttou{n_1}\ldots \inttou{n_m})\\
  &\leadsto^*& \inttou{f(g_1(n_1,\ldots,n_m),\ldots,g_n(n_1,\ldots,n_m))}.
\end{eqnarray*}
Now, suppose $f:\N^m\rightarrow\N$ and $g:\N^{m+2}\rightarrow\N$ and let
$h:\N^{m+1}\rightarrow\N$ be the function obtained by $f$ and $g$ by
primitive recursion, i.e.
\begin{eqnarray*}
h(0,n_1,\ldots,n_m)&=&f(n_1,\ldots,n_m);\\
h(n+1,n_1,\ldots,n_m)&=&g(n,h(n,n_1,\ldots,n_m),n_1,\ldots,n_m).
\end{eqnarray*}
We define
\begin{eqnarray*}
N&\equiv& \lambda x_m.\lambda y_m.\ldots.\lambda x_1.\lambda y_1.\lambda y.\lambda w.M_g y(w x_1\ldots x_m)y_1\ldots y_m;\\
M_h^m&\equiv&[N x_my_m]_{x_m,y_m}^{z_n};\\
\forall i<m.M_h^i&\equiv&[\lambda z_{i+1}.M_h^{i+1} x_i y_i]_{x_i,y_i}^{z_i};\\
N_h&\equiv&\lambda y.\lambda w.\lambda z_1.\ldots.\lambda z_m.M_h^1z_2\ldots z_n yw;\\
M_h&\equiv&\lambda x.x\llan N_h,M_f\rran.
\end{eqnarray*}
Notice that:
\begin{eqnarray*}
\inttou{0}\llan N_h,M_f\rran&\leadsto& M_f\equiv V_0;\\
\inttou{n+1}\llan N_h,M_f\rran&\leadsto& N_h \inttou{n} (\inttou{n}\llan N_h,M_f\rran)\\
  &\leadsto^*& \lambda z_1.\ldots.\lambda z_m.M_h^1z_2\ldots z_n \inttou{n}(\inttou{n}\llan N_h,M_f\rran)\equiv V_n.
\end{eqnarray*}
and moreover, for every $\inttou{n_1},\ldots,\inttou{n_m}$:
\begin{eqnarray*}
V_0\inttou{n_1}\ldots\inttou{n_m}&\leadsto^*&M_f\inttou{n_1}\ldots\inttou{n_m}\leadsto\inttou{h(0,n_1,\ldots,n_m)};\\
V_{n+1}\inttou{n_1}\ldots\inttou{n_m}&\equiv&
  (\lambda z_1.\ldots.\lambda z_m.M_h^1z_2\ldots z_n \inttou{n}(\inttou{n}\llan N_h,M_f\rran))\inttou{n_1}\ldots\inttou{n_m}\\
 &\leadsto^*& M_h^1\{\inttou{n_1},z_1\}\inttou{n_2} \ldots \inttou{n_m} \inttou{n}(\inttou{n}\llan N_h,M_f\rran)\\
 &\leadsto^*& (\lambda z_{2}.M_h^{2} \inttou{n_1}\inttou{n_1})\inttou{n_2} \ldots \inttou{n_m} \inttou{n}
  (\inttou{n}\llan N_h,M_f\rran)\\
 &\leadsto^*& (\lambda z_{3}.M_h^{3} \inttou{n_2}\inttou{n_2}\inttou{n_1}\inttou{n_1})\inttou{n_3} \ldots 
  \inttou{n}(\inttou{n}\llan N_h,M_f\rran)\\
 &\leadsto^*&\ldots\\
 &\leadsto^*& N \inttou{n_m}\inttou{n_m}\ldots \inttou{n_1}\inttou{n_1}\inttou{n}(\inttou{n}\llan N_h,M_f\rran)\\
 &\leadsto^*& M_g\inttou{n} (\inttou{n}\llan N_h,M_f\rran \inttou{n_1}\ldots \inttou{n_m}) \inttou{n_1}\ldots\inttou{n_m}\\
 &\leadsto^*& M_g\inttou{n} (V_n\inttou{n_1}\ldots \inttou{n_m}) \inttou{n_1}\ldots\inttou{n_m}\\  
 &\leadsto^*& M_g\inttou{n}\inttou{h(n,n_1,\ldots,n_m)} \inttou{n_1}\ldots\inttou{n_m}\\  
 &\leadsto^*& \inttou{g(n,h(n,n_1,\ldots,n_m),n_1,\ldots,n_m)}\equiv\inttou{h(n+1,n_1,\ldots,n_m)}.
\end{eqnarray*}
This concludes the proof.\qed
\end{proof}
The following two results show that functions
representable in $\rh{\emptyset}$ (respectively,
$\rh{\mathsf{A}}$) combinatorially saturate
$\mathbf{FP}$ (respectively, $\mathbf{FE}$).
\begin{lemma}\label{lemma:coeaddsqu}
There are terms $\Coerc,\Add,\Square$ such that for every $n,m$
$$
\begin{array}{rcl}
\Coerc\;\inttou{n}&\leadsto^*&\inttou{n};\\
\Add\;\inttou{n}\;\inttou{m}&\leadsto^*&\inttou{n+m};\\
\Square\;\inttou{n}&\leadsto^*&\inttou{n^2}.
\end{array}
$$
Moreover, for every $i\in\N$
\begin{eqnarray*}
\vdash_{\rh{\emptyset}}\Coerc&:&\U^{i+1}\multimap\U^i;\\
\vdash_{\rh{\emptyset}}\Add&:&\U^{i+1}\multimap\U^i\multimap\U^i;\\
\vdash_{\rh{\emptyset}}\Square&:&\U^{i+2}\multimap\U^i.
\end{eqnarray*}
\end{lemma}
\begin{proof}
$\Coerc$ is $\lambda x.x\llan\lambda y.\lambda w.c_1^\U w,c_{2}^\U\rran$. 
$\Add$ is $\lambda x.\lambda y.(x\llan M_1,M_2\rran)y$,
where $M_1$ is $\lambda w.\lambda z.\lambda q.c^\U_1(z q)$ and
$M_2$ is $\lambda z.z$. $\Square$ is 
$$
\lambda x.[\Add((\Coerc\;x_1)\llan\Add,c_2^\U,\rran)((\Predecessor\;x_2)\llan\Add,c_2^\U\rran)]_{x_1,x_2}^x
$$
where $\Predecessor$ is $\lambda x.x\llcu\lambda y.y,c_2^\U \rrcu$. Indeed:
\begin{eqnarray*}
c_2^\U\llan\lambda y.c_1^\U y,c_{2}^\U\rran&\leadsto& c_2^\U;\\
c_1^\U t\llan\lambda y.\lambda w.c_1^\U w,c_{2}^\U\rran
 &\leadsto&(\lambda y.\lambda w.c_1^\U w)t(t\llan\lambda y.\lambda w.c_1^\U w,c_{2}^\U\rran)\\
 &\leadsto^*&(\lambda y.\lambda w.c_1^\U w)tt\\
 &\leadsto^*&c_1^\U t;\\
\Coerc\;t&\leadsto&t\llan\lambda y.\lambda w.c_1^\U w,c_{2}^\U\rran\leadsto^*t;\\
\inttou{0}\llan M_1,M_2\rran&\leadsto&\lambda y.y\equiv V_0;\\
\inttou{n+1}\llan M_1,M_2\rran&\leadsto&M_1\inttou{n}(\inttou{n}\llan M_1,M_2\rran)\\
 &\leadsto&(\lambda z.\lambda q.c_1^\U(z q))V_n\leadsto \lambda q.c_1^\U(V_n q)\equiv V_{n+1};\\
V_0\inttou{m}&\leadsto&\inttou{m};\\
V_{n+1}\inttou{m}&\leadsto& c_1^\U(V_n \inttou{m})\leadsto^* c_1^\U\inttou{n+m}\equiv\inttou{(n+1)+m};\\
\Add \inttou{n}\inttou{m}&\leadsto&(\inttou{n}\llan M_1,M_2\rran)\inttou{m}\\
 &\leadsto^*& V_{n}\inttou{m}\leadsto^*\inttou{n+m};\\
\inttou{0}\llan \Add,c_2^\U\rran&\leadsto&\inttou{0}\equiv\inttou{0(0+1)/2};\\
\inttou{n+1}\llan \Add,c_2^\U\rran&\leadsto&\Add\inttou{n}(\inttou{n}\llan \Add,c_2^\U\rran)\\
 &\leadsto&\Add\inttou{n}(\inttou{n(n+1)/2})\leadsto^*\inttou{n+n(n+1)/2}\equiv\inttou{(n+1)(n+2)/2};\\
\Square\inttou{0}&\leadsto^*&\Add((\Coerc \inttou{0})\llan\Add,c_2^\U,\rran)((\Predecessor \inttou{0})\llan\Add,c_2^\U\rran)\\
 &\leadsto^*&\Add(\inttou{0}\llan\Add,c_2^\U,\rran)(\inttou{0}\llan\Add,c_2^\U\rran)\leadsto^*\Add\inttou{0}\inttou{0}\\
 &\leadsto^*&\inttou{0};\\
\Square\inttou{n+1}&\leadsto^*&\Add((\Coerc \inttou{n+1})\llan\Add,c_2^\U,\rran)\\
 &&((\Predecessor \inttou{n+1})\llan\Add,c_2^\U\rran)\\
 &\leadsto^*&\Add(\inttou{n+1}\llan\Add,c_2^\U,\rran)(\inttou{n}\llan\Add,c_2^\U\rran)\\
 &\leadsto^*&\Add\inttou{(n+1)(n+2)/2}\inttou{n(n+1)/2}\\
&\leadsto^*&\inttou{(n+1)(n+2)/2+n(n+1)/2}\equiv\inttou{n^2}.
\end{eqnarray*}
This concludes the proof.\qed
\end{proof}
In presence of ramification, an exponential behavior can
be obtained by exploiting contraction on tree-algebraic
types:
\begin{lemma}\label{lemma:exp}
There is a term $\Exp$ such that for every $n$
$$
\Exp\;\inttou{n}\leadsto^*\inttou{2^n}.
$$
Moreover, for every $i\in\N$
$$
\vdash_{\rh{\mathsf{A}}}\Exp:\U^{i+2}\multimap\U^i.\\
$$
\end{lemma}
\begin{proof}
For every $n\in\N$, we will denote by $\mathit{ct}(n)$ the 
complete binary tree of height $n$ in $\mathscr{E}_\C$:
\begin{eqnarray*}
\mathit{ct}(0)&=&c_2^\C;\\
\mathit{ct}(n+1)&=&c_1^\C(\mathit{ct}(n))(\mathit{ct}(n)).
\end{eqnarray*}
For every $n$, there are $2^n$ instances of 
$c_2^\C$ inside $\mathit{ct}(n)$. We will now define
two terms $\Blowup$ and $\Leaves$ such that
\begin{eqnarray*}
\Blowup\;\inttou{n}&\leadsto^*&\mathit{ct}(n);\\
\Leaves\;(\mathit{ct}(n))&\leadsto^*&\inttou{2^n}.
\end{eqnarray*}
We define:
\begin{eqnarray*}
\Blowup&\equiv&\lambda x.x\llan\lambda y.\lambda w.c_1^\C ww,c_2^\C\rran;\\
\Leaves&\equiv&\lambda x.(x\llan\lambda y.\lambda w.\lambda z.\lambda q.\lambda r.z(qr).\lambda x.c_1^\U x\rran)c_2^\U;\\
\Exp&\equiv&\lambda x.\Leaves(\Blowup x).
\end{eqnarray*}
Indeed:
\begin{eqnarray*}
\inttou{0}\llan\lambda y.c_1^\C yy,c_2^\C\rran&\leadsto&c_1^\C\equiv \mathit{ct}(0);\\
\inttou{n+1}\llan\lambda y.\lambda w.c_1^\C ww,c_2^\C\rran&\leadsto&
  (\lambda y.\lambda w.c_1^\C ww,c_2^\C)\inttou{n}(\inttou{n}\llan\lambda y.\lambda w.c_1^\C ww,c_2^\C\rran)\\
 &\leadsto^*&(\lambda y.\lambda w.c_1^\C ww,c_2^\C)\inttou{n}(\mathit{ct}(n));\\
 &\leadsto^*&c_1^\C(\mathit{ct}(n))(\mathit{ct}(n))\equiv\mathit{ct}(n+1)\\
\Blowup\inttou{n}&\leadsto&\inttou{n}\llan\lambda y.c_1^\C yy,c_2^\C\rran\leadsto(\mathit{ct}(n));\\
(\mathit{ct}(0))\llan\lambda y.\lambda w.\lambda z.\lambda q.\lambda r.z(qr).\lambda x.c_1^\U x\rran&\leadsto&
  \lambda x.c_1^\U x\equiv V_0;\\
(\mathit{ct}(n+1))\llan\lambda y.\lambda w.\lambda z.\lambda q.\lambda r.z(qr).\lambda x.c_1^\U x\rran&\leadsto^*&
 \lambda r.V_n(V_n r)\equiv V_{n+1};\\
V_0\inttou{m}&\leadsto&\inttou{1+m}\equiv\inttou{2^0+m};\\
V_{n+1}\inttou{m}&\leadsto& V_n(V_n\inttou{m})\leadsto^*V_n\inttou{2^n+m}\\
  &\leadsto^*&\inttou{2^n+2^n+m}\equiv\inttou{2^{n+1}+m};\\
\Leaves(\mathit{ct}(n))&\leadsto&
  ((\mathit{ct}(n))\llan\lambda y.\lambda w.\lambda z.\lambda q.\lambda r.z(qr).\lambda x.c_1^\U x\rran)c_2^\U\\
 &\leadsto^*& V_n\inttou{0}\leadsto^*\inttou{2^n};\\
\Exp\inttou{n}&\leadsto&\lambda x.\Leaves(\Blowup \inttou{n})\leadsto^*\Leaves(\mathit{ct}(n))\\
 &\leadsto^*&\inttou{2^n}.
\end{eqnarray*}
This concludes the proof.\qed
\end{proof}
The last two lemmata are not completeness results, but help
in the so-called \emph{quantitative} part of the encoding of Turing Machines. Indeed,
$\mathbf{FP}$ can be embedded into $\rh{\emptyset}$, while $\mathbf{FE}$ can
be embedded into $\rh{\mathsf{A}}$: 
\begin{theorem}
For every polynomial time computable function $f:\{0,1\}^*\rightarrow\{0,1\}^*$ there
are a term $M_f$ and an integer $n_f$ such that $\vdash_{\rh{\emptyset}} M_f:\B^{n_f}\rightarrow\B^0$
and $M_f$ represents $f$. For every elementary time computable function $f:\{0,1\}^*\rightarrow\{0,1\}^*$ 
there are a term $M_f$ and an integer $n_f$ such that $\vdash_{\rh{\mathsf{A}}} M_f:\B^{n_f}\rightarrow\B^0$
and $M_f$ represents $f$.
\end{theorem}
\begin{proof}
First of all, we can observe that for ever polynomial $p:\N\rightarrow\N$, there are
another polynomial $\overline{p}:\N\rightarrow\N$, an integer $n_p$ and term $M_{\overline{p}}$ such
that
\begin{eqnarray*}
\forall n\in\N.\overline{p}(n)&\geq&p(n);\\
\forall n\in\N.&\vdash_{\rh{\emptyset}}&M_{\overline{p}}:\B^{n_p+n}\linear\B^n;
\end{eqnarray*} 
and $M_{\overline{p}}$ represents $\overline{p}$. $\overline{p}$ is simply $p$
where all monomials $x^k$ are replaced by $x^{2^l}$ (where $k\leq 2^l$)
and $M_{\overline{p}}$ is built up from terms in $\mathscr{E}_\U$,
$\Add$, $\Square$ and $\Coerc$ (see 
Lemma~\ref{lemma:coeaddsqu}). Analogously, for every elementary function $p:\N\rightarrow\N$,
there are another elementary function 
$\overline{p}:\N\rightarrow\N$, an integer $n_p$ and term $M_{\overline{p}}$
such that 
\begin{eqnarray*}
\forall n\in\N.\overline{p}(n)&\geq&p(n);\\
\forall n\in\N.&\vdash_{\rh{\mathsf{A}}}&M_{\overline{p}}:\B^{n_p+n}\linear\B^n.
\end{eqnarray*} 
and $M_{\overline{p}}$ represents $\overline{p}$. This time, $\overline{p}$ 
is a tower 
$$
\overline{p}(n)=
2^{\textstyle \cdot^{\textstyle \cdot^{\textstyle \cdot^{\textstyle
2^n}}}}
\hskip 0.00cm\vbox{\hbox{$\Big\}\scriptstyle k\;\;\mathit{times}$}\kern0pt}
$$
obtained from $p$ by applying a classical result on elementary functions,
while $M_{\overline{p}}$ built up from terms in $\mathscr{E}_\U$, $\Add$, $\Coerc$ and $\Exp$ (see
Lemma~\ref{lemma:exp}).\par
Now, consider a Turing Machine $\mathcal{M}$ working in polynomial time. Configurations
for $\mathcal{M}$ are quadruples $(\mathit{state},\mathit{left},\mathit{right},\mathit{current})$,
where $\mathit{state}$ belongs to a finite set of states, $\mathit{left},\mathit{right}\in\Sigma^*$
(where $\Sigma$ is a finite alphabet) are the contents of the left and right portion
of the tape, and $\mathit{current}\in\Sigma$ is the symbol currently read by the head.
It it not difficult to encode configurations of $\mathcal{M}$ by terms
in $\mathscr{E}_\D$ in such a way that terms $M_\mathit{init},M_\mathit{final},M_\mathit{trans}$
exists such that:
\begin{varitemize}
  \item
  $M_\mathit{init}\bintob{s}$ rewrites to the term encoding the initial
  configuration on $s$, $M_\mathit{final}$ extract the result from a final configuration
  and $M_\mathit{trans}$ represents the transition function of $\mathcal{M}$;
  \item
  For every $n$,
  \begin{eqnarray*}
  \vdash_{\rh{\emptyset}}M_\mathit{init}&:&\B^{n+1}\linear\D^n;\\ 
  \vdash_{\rh{\emptyset}}M_\mathit{final}&:&\D^{n+1}\linear\B^n;\\
  \vdash_{\rh{\emptyset}}M_\mathit{trans}&:&\D^n\linear\D^n.
  \end{eqnarray*}
\end{varitemize}
Moreover, there is a term $M_\mathit{length}$ such that
$M_\mathit{length}\bintob{s}\leadsto^*\inttou{|s|}$
for every $s\in\{0,1\}^*$. Let now $p:\N\rightarrow\N$ be a polynomial
bounding the running time of $\mathcal{M}$. The function computed by $\mathcal{M}$
is the one represented by the term:
$$
M_\mathcal{M}\equiv\lambda x.\langle M_\mathit{final}(((M_{\overline{p}}(M_\mathit{length}y))
\llan\lambda x.\lambda y.\lambda w.y(M_\mathit{trans}w),\lambda x.x\rran)(M_\mathit{init}z))\rangle_{y,z}^x
$$
where $\langle M\rangle_{y,z}^x$ is the generalization of $[M]_{x,y}^z$ to the algebra $\B$.\par
If $\mathcal{M}$ works in elementary time, we can proceed in the same way.\qed
\end{proof}

\section{Conclusions}
We introduced a typed lambda-calculus equivalent to G\"odel
System $\mathsf{T}$ and a new context-based semantics
for it. We then characterized the 
expressive power of various subsystems of the
calculus, all of them being obtained by imposing linearity and
ramification constraints. To the author's
knowledge, the only fragment whose 
expressive power has been previously
characterized is $\rh{\mathsf{W}}$ 
(see~\cite{hofmann00safe,Bellantoni00apal,dallago03types}).
In studying the combinatorial dynamics of
normalization, the semantics has been exploited
in an innovative way.\par
There are other systems to which our semantics 
can be applied. This, in particular,
includes non-size-increasing polynomial time 
computation~\cite{Hofmann99lics} and
the calculus capturing $\mathbf{NC}$
by Aehlig et al.~\cite{Aehlig01ptcs}.  
Moreover, we believe higher-order contraction can
be accomodated in the framework by 
techiques similar to the ones 
from~\cite{Gonthier92popl}.\par
The most interesting development, however,
consists in studying the applicability of
our semantics to the automatic extraction
of runtime bounds from programs. This, however,
goes beyond the scope of this paper and is left
to future investigations.
\subsection*{Acknowledgements}
The author would like to thank Simone Martini and Luca Roversi for
their support and the anonymous referees for many useful comments.

\bibliographystyle{latex8}

\end{document}